\documentclass[12pt, referee]{aastex}
\usepackage[dvips]{color}
\newcommand{\re}{\textcolor[rgb]{0,0,0}}

\shorttitle{621 GHz water maser}
\shortauthors{Neufeld et al.}

\begin{document}

\title{Herschel$^*$/HIFI observations of a new interstellar water maser: the $5_{32}-4_{41}$ transition at 620.701~GHz}
\author{David A.~Neufeld\altaffilmark{1}, Yuanwei Wu\altaffilmark{2}, Alex Kraus\altaffilmark{2}, Karl M.\ Menten\altaffilmark{2}, Volker Tolls\altaffilmark{3}, Gary~J.~Melnick\altaffilmark{3}, and Zsofia~Nagy\altaffilmark{4}}

\altaffiltext{*}{Herschel is an ESA space observatory with science instruments provided
by European-led Principal Investigator consortia and with important
participation from NASA}
\altaffiltext{1}{Department\ of Physics \& Astronomy, Johns Hopkins University\,
3400~North~Charles~Street, Baltimore, MD 21218, USA}
\altaffiltext{2}{Max-Planck-Institut f\"ur Radioastronomie, Auf dem H\"ugel 69, 53121 Bonn, Germany}
\altaffiltext{3}{Harvard-Smithsonian Center for Astrophysics, 60 Garden Street, Cambridge, MA 02138, USA}
\altaffiltext{4}{Kapteyn Astronomical Institute, Landleven 12, 9747 AD, Groningen, The Netherlands}

\begin{abstract}

Using the {\it Herschel Space Observatory}'s Heterodyne Instrument for the Far-Infrared (HIFI), we have performed mapping observations of the 620.701~GHz $5_{32}-4_{41}$ transition of ortho-H$_2$O within a $\sim 1.5^\prime \times 1.5^\prime$ region encompassing the Kleinmann-Low nebula in Orion, and pointed observations of that transition toward the Orion South condensation and the W49N region of high-mass star formation.  Using the Effelsberg 100 m radio telescope, we obtained ancillary observations of the { 22.23508}~GHz $6_{16}-5_{23}$ water maser transition; in the case of Orion-KL, the 621~GHz and 22~GHz observations were carried out {within 10 days of each other}.   The 621~GHz water line emission shows clear evidence for strong maser amplication in all three sources, exhibiting narrow ($\sim 1$ km/s FWHM) emission features that are coincident ({kinematically} and/or spatially) with observed 22~GHz features.  \re{Moreover, in the case of W49N -- for which observations were available at three epochs spanning a two year period -- the spectra exhibited variability.}   The observed 621~GHz/22~GHz line ratios are consistent with a maser pumping model in which the population inversions arise from the combined effects of collisional excitation and spontaneous radiative decay, and the inferred physical conditions can plausibly arise in gas heated by either dissociative or non-dissociative shocks.  The collisional excitation model also predicts that the 
22~GHz population inversion will be quenched at higher densities than that of the 621~GHz transition, providing a natural explanation for the observational fact that 22~GHz maser emission appears to be a necessary but insufficient condition for 621~GHz maser emission.

\end{abstract}


\section{Introduction}

Barely fifteen years after the invention of the laboratory maser in 1953 -- which required a population inversion in the ammonia molecule to be carefully {\it engineered} -- interstellar masers were {\it discovered} as a remarkable and naturally-occurring phenomenon. The $6_{16}-5_{23}$ transition of water vapor, with a frequency near 22 GHz, was among the first masing transitions detected from an astrophysical source (the Orion Molecular Cloud; Cheung et al.\ 1969), and has proven to be an extraordinarily useful probe in environments as diverse as interstellar shock waves, circumnuclear gas in active galactic nuclei (AGN), and the envelopes of evolved stars (\re{Elitzur 1992; Lo 2005; and references therein}). With brightness temperatures that often exceed $10^{10}$~K and in extreme cases can 
\re{exceed $10^{15}$~K (e.g. Garay, Moran \& Haschick (1989)}, and with linewidths that are often extremely narrow, 22 GHz water masers can be observed with Very Long Baseline Interferometry  (VLBI).  {The very high angular resolution of 
such VLBI observations enables proper motion studies of the kinematics 
of gas outflowing from proto- and evolved stars.  Moreover, trigonometric 
and geometric/rotational parallaxes can be determined, 
yielding the distances and motions of star forming regions in our Galaxy, in Local Group galaxies, 
and in more distant AGN (e.g.\ Brunthaler et al.\ 2007; Braatz et al.\ 2010).  Such observations have led to revised estimates of
the size, shape and kinematics of the Milky Way 
(e.g.\ Reid et al.\ 2009),} as well as the best evidence yet obtained for the existence of supermassive black holes (e.g. Miyoshi et al.\ 1995).  Masers directly probe extreme environments (of density and temperature), whose properties can often only be inferred indirectly from other lines.

Over the 40 years since the first detection of interstellar water maser emission in the 22 GHz transition, several additional interstellar maser transitions have been detected at higher frequencies, some in response to a specific prediction of a population inversion (Menten et al. 1990a). In the case of pure rotational emissions, these comprise the 183 GHz (Waters et al. 1980), 321 GHz (Menten et al. 1990b), 325 GHz (Menten et al. 1990a), 380 GHz (Phillips et al. 1980), 439 GHz (Melnick et al. 1993) and 471 GHz (Melnick et al. 1993) lines, associated respectively with the $3_{13}-2_{20}$, $10_{29}-9_{36}$, $5_{15}-4_{22}$, $4_{14}-3_{21}$, $6_{43}-5_{50}$, and $6_{42}-5_{51}$ transitions.  These transitions are shown in the energy level diagram in Figure 1.  In addition, several water transitions within the the $\nu_2=$1 {and 2} vibrationally-excited states have been observed toward the circumstellar envelopes of evolved stars, {as have the 437 GHz $7_{53}-6_{60}$ and 475~GHz 5$_{33}-4_{41}$ pure rotational transitions (Melnick et al.\ 1993; Menten et al.\ 2008).}

The detection of multiple masing transitions has been crucial in elucidating the pumping mechanism responsible for the population inversions and the physical conditions in the masing gas. Understanding the latter, of course, is crucial to the astrophysical interpretation of maser observations. While the interpretation of the emission observed in a single line (e.g. the 22 GHz transition) is very geometry dependent, because the non-linear amplification of the radiation strongly favors those sight-lines that happen to possess the greatest velocity coherence, the interpretation of multiline observations is more
robust and permits important constraints to be placed upon (1) the excitation mechanism and (2) the conditions of temperature and density in the emitting gas. The pattern of pure rotational maser transitions observed from interstellar sources appears to confirm an excitation model (e.g. Neufeld \& Melnick 1991) in which collisional pumping, combined with spontaneous radiative decay, leads to the inversion of exactly those transitions that are observed to mase. In evolved stars, however, the additional presence of the 437 GHz $7_{53}-6_{60}$ maser transition suggests that collisional pumping is not the entire story and that radiative pumping by dust continuum radiation is also important. 

Quantitative measurements of maser line ratios provide additional constraints.  For example, the predicted dependence of a maser line ratio upon the temperature and density of the masing gas was used by Melnick et al. (1993) to derive a lower limit of 1000 K upon the gas temperature in several observed sources; this value argued against a model (Elitzur, Hollenbach \& McKee 1989) in which the masers were excited in 400 K material behind a dissociative shock wave, and favored a model (Kaufman \& Neufeld 1996) in which non-dissociative magnetohydrodynamic \re{shocks} were the source of the emission. The analysis makes the assumption that the maser beam angle is similar for the two transitions that are being compared. 

Prior to the launch of the {\it Herschel Space Observatory} (Pilbratt et al.\ 2010), the observational data on pure rotational water masers had been limited to transitions of frequency less than 500 GHz, largely because of atmospheric absorption. The HIFI spectrometer on {\it Herschel} (de Grauuw et al.\ 2010), however, provides the opportunity of expanding the available data set by the addition of higher frequency transitions that promise to increase our leverage on the pumping mechanism and the physical conditions in the maser-emitting gas.  To date, {\it Herschel} observations of oxygen-rich evolved stars have led to the detection of two additional water maser transitions: \re{(1) the $5_{32}-4_{41}$ transition at 620.701~GHz\footnote{\re{referred to hereafter as the ``621~GHz transition"}} (detected toward VY CMa by Harwit et al.\ 2010, and toward W Hya, IK Tau and IRC+10011 by Justtanont et al.\ 2012); and (2) the $5_{24}-4_{31}$ transition at 970.315~GHz (detected toward W Hya and IK Tau by Justtanont et al.\ 2012). To our knowledge, neither maser has previously been observed from interstellar gas.} 
In this paper, we report the first detection of 621~GHz water maser emission from {\it interstellar} gas, obtained with {\it Herschel}/HIFI in mapping observations of the Orion-KL region.  In addition, we report the detection of 621~GHz water maser emission obtained serendipitously in 
single pointings toward the W49N star-forming region and the Orion South molecular cloud. 

\section{Observations and data reduction}

Our {\it Herschel} observations of Orion-KL were carried out on 2011 March 11 as part of the ``Orion Small Maps'' subprogram within the HEXOS Guaranteed Time Key Program (GTKP; P.I., E.~Bergin).  We used the ``Heterodyne Instrument for the Infrared'' (HIFI), in ``on-the-fly mapping'' (OTF) mode, to obtain a Nyquist-sampled map consisting of a 6 by 8 rectangular array of pointings spaced by $\sim 16^{\prime\prime}$ in R.A.\ and declination.  The map center was located at offset ($\rm \Delta \alpha cos\delta, \Delta \delta)= (+1.5^{\prime\prime},+10.5^{\prime\prime})$ relative to Orion-KL.  (All offsets given in this paper are relative to $\rm 5h\,35m\,14.3s, -5d\,22^\prime\,33.7^{\prime \prime}$ (J2000), the position we adopt for Orion-KL.)
The reference position for the OTF mapping observations, at offset $(+841.5^{\prime\prime},+870.5^{\prime\prime})$, was chosen to be devoid of known molecular emission.  The observations were carried out in the upper sideband of mixer band 1b, using the WBS spectrometer, \re{which provides an oversampled channel spacing of 0.5 MHz (0.27 km/s at a frequency of 621 GHz), roughly one-half the effective resolution. The absolute frequency calibration is accurate to 100~kHz (Roefsema et al.\ 2012)}.  The beam size was 34$^{\prime \prime}$ (HPBW), \re{and the absolute pointing accuracy is $2^{\prime \prime}$}.  

Two concatenated Astronomical Observation Requests (AORs) were used to obtain two separate maps, with a small relative offset ($6.6^{\prime\prime}$)  chosen to make the center of the H polarization beam for one map coincident with the center of the V polarization beam for the other map.  As discussed in \S3, the goal of acquiring two separate maps in this manner was to obtain a measurement of any linear polarization in the maser feature.  As summarized in Table 1, the total duration of each AOR was 1987~s, with an on-source integration time of 420~s, which yielded individual spectra with an r.m.s. noise of 62 mK  (on the scale of antenna temperature and for a 1.1 MHz channel width).
The {\it Herschel} data on Orion-KL were reduced using standard methods in the Herschel Interactive Processing Environment (HIPE; Ott 2010), {version 10.0} 

{On 2011 March 21, ten days after 
the {\it Herschel} observations of Orion-KL were performed}, we used the Effelsberg 
100~m telescope to carry out observations of the 22.23508 GHz $6_{16}-5_{23}$ transition, with the goal of determining the 621 GHz / 22 GHz line flux ratio as a constraint upon models for the maser emission mechanism.  Here, we obtained a map, centered at offset $(+3.1^{\prime\prime}, +3.9^{\prime\prime})$, consisting of a 9 by 9 square array of pointings spaced by $\sim 20^{\prime\prime}$ in R.A.\ and declination.  The beam size was 41$^{\prime \prime}$ (HPBW), {and the frequency resolution was 6.1~kHz, corresponding to a velocity resolution of 0.082~km/s.  The observations were performed using the K-band receiver at the prime focus of the 100~m telescope, with an observing time of 1.0 min per position.}  In calibrating the spectra, we applied corrections for the atmospheric attenuation and for the dependence of the telescope gain on the elevation.  The calibration factor was determined by the observation of suitable calibration sources like NGC7027 and 3C286 (taking into account the significant linear polarization of the latter).

In addition to the mapping observations of Orion-KL that are the primary subject of this paper, we have also identified two additional {\it Herschel}/HIFI spectra that show narrow and/or time variable 621 GHz features suggestive of maser action.  Our observations of high-mass star-forming region W49N were carried out at three separate epochs, as part of the PRISMAS GTKP (P.I, M.~Gerin).  These observations had the primary goal of measuring foreground absorption in nearby spectral lines of $\rm H_2O^+$, but included the 621~GHz line within the bandpass.  For each epoch, the relevant data were acquired in 3 AORs obtained with slightly different LO settings.  The AOR numbers, dates of observation, beam center position, duration of the observations, and rms noise achieved are listed in Table 1.  The data reduction methods that we adopted  within the PRISMAS GTKP have been described, for example, by Neufeld et al.\ 2010.  Our observations of the {hot core in} the Orion South molecular cloud were performed as part of a spectral line survey in the HEXOS GTKP.  \re{The data of present interest were acquired in a full spectral scan of band 1b, in which the 621~GHz transition was observed (in either the upper or the lower sideband) at 19 separate LO settings.}  The data reduction methods that we adopted for spectral scans within the HEXOS GTKP have been described by Bergin et al.\ (2010), and further details of the observations appear in Table 1. 

Ancillary 22 GHz maser observations were performed using the Effelsberg 100 m telescope toward W49N and Orion S on 2012 Oct 1 and 2012 Sep 28, respectively.  Because water masers are well-known to exhibit {significant} variability on timescales {of months}, our comparison of these non-contemporaneous 22 GHz and 621 GHz spectra must be interpreted with caution; indeed, as described in \S3 below, time variability in the maser emission is readily inferred from a comparison of the W49N 621 GHz data acquired at the 3 epochs.

\section{Results}

\subsection{Mapping observations of Orion-KL}

In Figure 2, we show the 621 GHz $5_{32}-4_{41}$ line spectra obtained toward Orion-KL.   The spectra are autoscaled, with the color of each border indicating the vertical scaling in accordance with the color bars on the left, and the two linear polarizations are shown in separate panels.  The LSR velocity range shown in each panel is --20 to +30 km/s. 
Analogous results are shown for the 22 GHz $6_{16}-5_{23}$ transition in Figure 3.  Clearly, the strongest 621~GHz line emission originates in the vicinity of Orion-KL, and exhibits a broad profile characteristic of high-lying (and non-masing) rotational transitions of water observed by {\it Herschel}/HIFI (Melnick et al.\ 2010).  However, to the northwest of Orion-KL, close to the shocked material associated with Orion H$_2$ Peak 1 (Beckwith et al.\ 1978), a narrow emission feature is clearly present.  This feature is most apparent in the four spectra obtained closest to offset $(-15^{\prime\prime},+43^{\prime\prime})$, the average of which is shown in Figure 4 (e.g. upper panel, blue histogram).
The 22 GHz maser emission, by contrast, is dominated by multiple narrow emission components, the strongest lying in the 7 to 13 km/s LSR velocity range; see Figure 4 (upper panel, red histogram).  
In the lower panel of Figure 4, we compare the 621 GHz and 22 GHz line emissions observed in the vicinity of $(-15^{\prime\prime},+43^{\prime\prime})$ for the 9 to 14 km/s LSR velocity range.  In this panel, the two polarizations are shown separately for each transition, and a broad emission component has been subtracted from the 621 GHz line profile.  Although the 621~GHz and 22~GHz spectra are dissimilar outside this velocity range -- the 22~GHz spectrum exhibiting a strong narrow feature near $v_{\rm LSR} \sim 7.5$~km/s that is absent in the 621~GHz spectrum, for example -- the spectra are very similar within the 10 -- 13 km/s velocity range.  \footnote{In Figure 4, the 22~GHz rest frequency is taken as 22.23508 GHz (Kukolich 1969), the average over all hyperfine components.  The three strongest hyperfine components (for which $\Delta F = -1$ and which are expected to account for more than $99.8\%$ of the observed emission,) lie at frequency shifts of $-36$, $-3$ and $+40$ kHz relative to the mean $6_{16}-5_{23}$ line frequency, corresponding to velocity shifts of $0.49$, $0.04$, and $-0.55\, \rm km/s$.
Thus, even if departures from LTE lead to large changes in the hyperfine line ratios, the maximum resulting velocity shift is at most 0.5 km/s; hyperfine splitting of the higher-frequency 621~GHz transitions is completely negligible.}
Furthermore, for both lines, the spectra obtained in the two polarizations are very similar.  Given the systematic uncertainties (associated, for example, with possible differences between the telescope beam profiles for the two polarizations), these data provide no compelling evidence for polarization.  \re{A more detailed study of polarization in this source, which makes use of
additional observations obtained at multiple spacecraft roll angles to
measure the linear polarization of the 621 GHz maser transition, is currently under
way (Jones et al.\ 2013, private communication)}

\subsection{Pointed observations toward W49N and Orion South}

The upper panel of Figure 5 shows the 22~GHz spectrum obtained toward W49N (red), along with the 621~GHz spectra obtained at three epochs spanning two years (see Table 1).  Because the 621 GHz transition was not specifically targeted but rather observed serendipitously in observations with a different primary purpose, we did not perform near-contemporaneous 22~GHz observations.  Thus, the 22 GHz spectrum shown in Figure 5 was acquired 170 days after the last 621~GHz observation, making a detailed comparison between the transitions difficult.  On the other hand, the availability of 621~GHz line observations at three separate epochs has allowed us to detect unequivocal variability in the line profile.  Most notably, a narrow emission feature near $v_{\rm LSR} = -59$~km/s was present in the 2011 observation but absent in the 2010 and 2012 observations; and the line profile in the $v_{\rm LSR} = 15-20$~km/s range shows clear variability throughout the period covered by the three epochs.  This variability, which can result from small changes in the (negative) optical depth along the sight-line to the observer, is characteristic of high-gain maser emission from a turbulent medium.  

\re{Four interferometric studies provide information about the spatial distribution of the 22~GHz water maser emission in W49N over a 30-year time period (Moran et al.\ 1973; Walker, Matsakis, \& Garcia-Barreto 1982; Gwinn, Moran \& Reid 1992; McGrath, Goss \& De~Pree 2004; performed in 1970--71, 1978, 1980--82, and 1998--99, respectively).  They indicate that 99\% of the 22~GHz maser flux is associated with the ultracompact HII regions designated G1 and G2 (Dreher et al.\ 1984), and originates within 5$^{\prime\prime}$ of the beam center position adopted for our {\it Herschel} (and Effelsberg) observations.   The remainder of the observed 22 GHz maser emission originates from a region $\sim 19^{\prime\prime}$ northeast of the beam center.  Given these offsets, the {\it Herschel} beam size (34$^{\prime\prime}$), and the absolute pointing accuracy (2$^{\prime\prime}$), we find it very unlikely that the observed 621~GHz variability could be the result of pointing errors.  The only way that pointing errors could falsely indicate variability is if a very strong 621 GHz maser emission feature were located in the far wing of the {\it Herschel} beam profile (where the beam response is a strong function of offset); such an emission feature would lie outside the region from which 22 GHz emission has been detected in the interferometric studies discussed above.}

\re{In the case of the 621~GHz spectra obtained toward W49N, the relative contribution of maser and non-maser emission is hard to disentangle, at least based on the data currently in hand.  While it is tempting to interpret the spectra as a superposition of narrow maser features on top of a broad ($\sim 40$~km/s wide) plateau of thermal line emission, the 22~GHz spectrum has an entirely  similar appearance.  In the case of the 22~GHz transition, however, the expected contribution of thermal emission is negligible; indeed, the interferometric studies referenced above show that the broad pedestal apparent in single-dish observations is - in reality - a superposition of numerous compact and narrow emission features within the beam.}

The lower panel of Figure 5 shows the spectra obtained toward Orion South in (non-contemponeous) observations of the 621~GHz and 22~GHz transitions.  The 621~GHz spectrum shows a single prominent narrow feature of width $\sim 1.5$~km/s (FWHM) in near coincidence with one of three narrow 22~GHz features.  The 621~GHz linewidth is considerably narrower than that ($\sim 4$~km/s) measured at this position for non-masing spectral lines  -- such as those of H$_2$Cl$^+$ (Neufeld et al.\ 2012) and CN (Rodriguez-Franco et al.\ 2001), for example -- and has its centroid at a lower LSR velocity ($\sim 6$~km/s versus of 7.2~km/s). {The strongest 22 GHz emission in Orion South was detected in to 0 to +17~km~s$^{-1}$ LSR velocity range, but weaker features are present over a wider interval from $-45$ to $+43$~km~s$^{-1}$; the latter range is somewhat wider than that obtained by Gaume et al.\ 1998 ($-20$ to $+44$~km~s$^{-1}$; see their Fig. 5 and Table 2) in sensitive Very Large Array observations performed in 1991.}  \re{The phenomenon of maser amplification in W49N and Orion S is not unique to water.  Maser emission from hydroxyl radicals has been long been studied in W49N, but is believed to require different excitation conditions (e.g. Mader et al.\ 1975).  In Orion South, Voronkov et al.\ (2005) have reported 6.7~GHz $\rm CH_3OH$ emissions that are likely masing.}

\section{Discussion}

\subsection{Observed line ratios}

\subsubsection{Spatially-averaged line ratios}

In Table 2, we present the 22~GHz and 621~GHz photon luminosities for the three sources we have observed, each computed with the assumption that the emission is isotropic and for adopted distances of 414~pc (Orion-KL and Orion-S; Menten et al.\ 2007) and 11400~pc (W49N: Gwinn, Moran \& Reid 1992).   As we have noted in \S3 above, however, particular velocity components that are clearly present in the 22~GHz spectra are sometimes unaccompanied by detectable 621~GHz line emission (though the converse is never true), suggesting that the requirements for strong 621~GHz maser amplification are more stringent than those for strong 22~GHz maser action.  Thus, the source-averaged 621~GHz/22~GHz line ratios may be considerable smaller than those applying to specific 621~GHz emission features.  Accordingly, we have also computed luminosities and line ratios for specific velocity ranges in which strong 621~GHz emission is observed from Orion-KL and Orion-S. Furthermore, as will be discussed further in \re{\S4.1.2} below, our mapping observations of Orion-KL indicate that {\it even when integrated over a narrow range of LSR velocities}, the 621~GHz / 22~GHz line ratio varies spatially.

\subsubsection{Spatial variations in Orion-KL}

In Figure 6, we present channel maps for the 22 GHz line emissions detected from Orion-KL, with the location of the peak intensity marked by a red diamond, and the channel-averaged peak flux (in Jy/beam) listed in red near the top of each panel.  The two strongest velocity components, at $v_{\rm LSR} \sim 7.5$~km/s and $v_{\rm LSR} \sim 11.8$~km/s show peak intensities near $(+4^{\prime\prime},-6^{\prime\prime})$ and $(+20^{\prime\prime},+14^{\prime\prime})$ respectively, positions separated by 26$^{\prime\prime}$, more than one-half the Effelberg HPBW (41$^{\prime\prime}$). \footnote{\re{The $\sim 7.5$~km/s maser emission feature can be identified with a bursting maser feature observed using VLBI by Hirota et al.\ (2011) and located at offset $(+2.6^{\prime\prime},-2.8^{\prime\prime})$.  This feature, which brightened dramatically in 2011 February (Tolmachev 2011), is associated the Orion Compact Ridge.  Hirota et al.\ suggested that the bursting maser emission arises in shocked gas associated with the interaction of an outflow with ambient material, and identified Radio Source I (Plambeck et al.\ 2009) as the likely origin of the outflow.  The latter is located roughly 3400 AU northeast of the bursting maser source, and is a site of well-studied SiO maser emission (e.g. Matthews et al.\ 2010).}}
Moreover, as shown in Figure 7 (yellow contours), the peak 22~GHz line intensity integrated over the 10 -- 13 km/s velocity range (upper panel) is clearly separated also from the location of the peak 621~GHz emission (lower panel) in the same velocity range, despite the similarity of the spectral line profiles shown in Figure 4 (lower panel).  One intepretation of this separation is that the 22~GHz maser emission in this velocity range is actually the sum of two (or more) spatially distinct components, the stronger of which is unaccompanied by 621 GHz maser emission.  

To test this hypothesis, we have carefully compared the 10 -- 13 km/s 22~GHz channel map with the beam profile for the Effelsberg telescope.  The latter was measured at 22 GHz by observing the compact continuum source 3C84 \re{at the prime focus}.  The beam response function is shown in Figure 8 (black contours, upper panel), superposed on red contours indicating the best fit Gaussian (with a HPBW of $41.03^{\prime\prime}$).  In each case, contours are labeled as a percentage of the peak response.  The beam profile is reasonably well-approximated by a Gaussian, with deviations less than $\sim 5\%$ of the peak intensity.  The lower panel shows a grayscale representation of the residuals relative to the best-fit Gaussian, with gray levels separated by 1$\%$ of the peak intensity and the blue contour indicating a residual of zero.  Having determined the best-fit to the 22~GHz beam profile, we subtracted that beam profile -- centered at the centroid of the strongest 22~GHz maser emission (at offset [$+15.7^{\prime \prime}$,$+13.6^{\prime \prime}$]) and rotated as appropriate for the mean parallactic angle of the Orion-KL observations (19$^0$)\footnote{\re{During our observations of Orion-KL, the parallactic angle varied over a relatively small range: from 7$^0$ to 27$^0$}}
-- from the observed profile of the 22~GHz maser emission for $v_{\rm LSR} = 10 - 13$~km/s.  The grayscale background in Figure 7 (both panels) represents the result of that subtraction, which we will henceforth refer to as the ``22 GHz residual" (i.e. the residual with respect to the intensity expected for a point source at offset [$+15.7^{\prime \prime}$,$+13.6^{\prime \prime}$]).  Here, adjacent grayscale levels are separated by 1000~K~km/s, corresponding to $\sim 5\%$ of the peak 22~GHz line intensity integrated over the 10 -- 13 km/s velocity range.  The spatial distribution of the ``22 GHz residual" matches that of the 621~GHz line emission reasonably well, supporting the interpretation introduced at the of end the previous paragraph.  Thus, the ratio of the luminosity of the 621~GHz line to that of the 22 GHz {\it residual} provides our best estimate of the true maser line ratio in the 621~GHz-emitting gas: 0.28 (Table 2).   
\re{It is difficult to provide a quantitative estimate of the systematic uncertainty in the 22 GHz residual; the latter depends upon the reproducibility of the Effelsberg beam profile shown in Figure 8, an issue that is not presently constrained by observations.  We note, however, that the peak flux in the residual is $\sim 20\%$ of the peak flux in the 10 -- 13 km/s velocity range. By comparison, the maximum deviations from a Gaussian in the beam response function (blue contours in Figure 8) are only 5$\%$ of the peak response.  This suggests that the likely error in our estimate of the residual flux is at most 25$\%$, even under the conservative assumption that the deviations from a Gaussian profile have locations that vary from one observation to the next.}

\re{In addition to comparing our observations of 621~GHz water emission with single-dish
22~GHz observations (performed nearly contemporaneously), we may also compare them with interferometric observations of the 22~GHz maser transition.  Unfortunately, we are unaware of any such observations of the Orion region performed in recent years.   The only published map of 22~GHz maser features to cover the region of narrow 621~GHz emission is that presented by Genzel et al.\ (1981), based upon observations performed in 1977--79; more recent observations reported by Gaume et al.\ 1998 do not extend far enough to the north.  We have, however, located a set of more recent unpublished data, acquired using Very Large Array (VLA) on 1996 March 15 and covering the region of interest, in the National Radio Astronomy Observatory (NRAO) archive.  Our reduction of these data is described in the Appendix, where we tabulate the velocities and positions of 339 maser spots of peak flux density exceeding 1 Jy that were identified in our analysis.  In Figure 7, blue, cyan and red circles show the locations of maser spots with LSR velocities in the 10 -- 11, 11 -- 12, and 12 -- 13~km/s ranges, with the size of each circle indicating the peak flux on a logarithmic scale.  Figure 7 indicates that a strong 22~GHz maser counterpart was indeed present at a location and velocity close to that of the narrow 621~GHz emission, despite the 15 year time period that separates the two observing epochs.}  

\subsection{Comparison with maser excitation models}

To help interpret the observed maser line ratios, we have updated the maser excitation models described by (Neufeld \& Melnick 1991; hereafter NM91) to make use of recent calculations of the rate coefficients for excitation of $\rm H_2O$ by H$_2$ (Daniel, Dubernet, \& Grosjean 2011, extrapolated for additional states of H$_2$O using the artificial neural network method introduced by Neufeld 2010).  As in NM91, we performed a parameter study in which the equations of statistical equilibrium for the H$_2$O level populations were solved for a grid of temperatures, $T$, H$_2$ densities, $n({\rm H}_2)$, and water column densities, $\rm N({\rm H_2O})$.  These calculations were performed for a medium with a large velocity gradient, with use of a standard escape probability method to account for radiative trapping.  These calculations yield two key quantities for any masing transition: \re{(1) the line optical depth in the unsaturated limit, and (2) the line luminosity in the saturated limit}.  

We computed the negative Sobolev optical depth, along the direction of the velocity gradient, that would \re{be obtained} without the effect of maser amplification.  These values are plotted in Figure 9, as a function of $n({\rm H_2})$ and $T$, for four values of $\rm N({\rm H_2O})$.  Here, red contours show results for the 22~GHz transition, and blue contours those for the 621~GHz transition.  For both transitions, the Sobolev optical depth is more negative than --3 within large regions of the parameter space plotted in Figure 9.  At high densities, the level populations inevitably tend to their local thermodynamic equilibrium (LTE) values, maser action is quenched, and the optical depths become positive (in regions to the right of the contours labeled zero.)  As expected, the lower-frequency 22~GHz transition can maintain its population inversion up to higher gas densities \re{than} the 621~GHz transition, allowing for the possibility of maser amplification of the 22~GHz transition unaccompanied by maser emission in the 621~GHz transition.  As discussed in \S 3, {\it this possibility is clearly realized in the observations reported here}. \re{For example, the strong 14.5~km/s 22~GHz feature in Orion South shows no detectable 621~GHz emission; the implied upper limit on the 621~GHz/22~GHz line ratio is $\sim 0.01$).  At high densities (e.g. $\sim 10^9 \rm \, cm^{-3}$ at a temperature of 1000~K), the maser line ratio depends strongly upon temperature and density: thus, relatively modest variations in the physical conditions can lead to large changes in the emergent line ratio.}

The optical depths plotted in Figure 9 are the values along sightlines {\it parallel} to the direction of the velocity gradient (i.e. in the direction of {\it minimum} velocity coherence.)  In plane-parallel geometry, the optical depth of a masing transition formally approaches $-\infty$ as the inclination of the ray approaches $90^0$, although in reality the magnitude of the optical depth is limited by departures from plane-parallel geometry (such as curvature in a shocked molecular shell, for example).  Nevertheless, the maximum optical depth in a typical interstellar region can easily exceed the Sobolev optical depth by an order of magnitude: thus, a Sobolev optical depth of 3 can easily produce a maser gain of $e^{30}$.  Gain factors of this magnitude will reduce the population inversion, leading to maser saturation. 

We computed the maximum photon luminosity that can \re{be obtained} under conditions of saturation, with the use of the formalism discussed in NM91. In essence, this is the maximum rate of stimulated emission that can occur without eliminating the population inversion.  Taking the ratio of these luminosities for the two transitions yields the results shown in Figure 10.  If the 22~GHz and 621~GHz transition are both saturated and are similarly beamed, this luminosity ratio can be compared with the observed values in rightmost column of Table 2.  This comparison indicates that the observed luminosity of the 621~GHz maser transition can be accounted for in the same collisional pumping model that has proved successful in explaining other masing transitions of interstellar water.  In no velocity range, in any source, does the 621~GHz/22~GHz maser luminosity ratio exceed 1.16; given the likelihood of time variability (in W49N and Orion South where the observations were non-contemporaneous) and other uncertainties associated with maser beaming, {\it this value can be regarded as consistent with the range of values (up to 0.8) predicted by the collisional excitation model}.  Unfortunately, and unlike some of the other submillimeter maser transitions (e.g.\ Melnick et al. 1993), the 621~GHz transition does not provide strong constraints on the gas temperature; thus the observed 621~GHz/22~GHz line ratios are consistent with either 400~K gas behind dissociative J-type shocks (e.g. Elitzur, Hollenbach \& McKee 1989), or the hotter gas that can arise behind non-dissociative shocks (Kaufman \& Neufeld 1993).  \re{A search for maser emission in the $6_{42}-5_{51}$ (471~GHz) and $6_{43}-5_{50}$ (439~GHz) transitions of water could provide a stronger constraint on the gas temperature, as these transitions can only be pumped significantly at temperatures $\sim 900$~K or higher (Melnick et al.\ 1993).} 

\section{Summary}

\noindent We have performed {\it Herschel}/HIFI  mapping observations of the 621~GHz $5_{32}-4_{41}$ transition of ortho-H$_2$O within a $\sim 1.5^\prime \times 1.5^\prime$ region encompassing the Kleinmann-Low nebula in Orion, and pointed observations of that transition toward the Orion South condensation and the W49N region of high-mass star formation. We also obtained ancillary observations of the {22.23508}~GHz $6_{16}-5_{23}$ water maser transition using the Effelsberg 100 m telescope.  In the case of Orion-KL, the 621~GHz and 22~GHz observations were carried out {nearly contemporaneously (within 10 days of each other)}.   The key results of these observations are as follows:
 
\noindent 1.  In all three sources, the 621~GHz water line emission shows clear evidence for strong maser amplification.  
In both Orion-KL and Orion South, the observations reveal a narrow ($\sim 1$ km/s FWHM) emission feature, superposed in the case of Orion-KL on broad thermal line emission, with a velocity centroid that is consistent with that of an observed 22~GHz maser feature.  In the case of W49N - for which observations were available at three epochs spanning a two year period - observed variability in the 621~GHz line profile provides further evidence for maser action.  

\noindent 2. In all three sources, 621~GHz maser emission is apparently always accompanied by 22~GHz maser emission.  Our mapping observations of Orion-KL reveal a spatial offset between the 621~GHz maser emission and the 22~GHz maser emission occuring at the same velocity.  However, a careful {analysis of the spatial distribution of the observed 22 GHz maser emission} reveals two spatially-distinct components, the weaker of which is spatially coincident with the 621~GHz maser emission.

\noindent 3. 22~GHz maser emission is not invariably accompanied by 621~GHz maser emission.  All three sources show some strong 22~GHz velocity components with no detectable 621~GHz counterpart.

\noindent 4.  The observed 621~GHz/22~GHz line ratios are always consistent with a maser pumping model in which the population inversions arise from the combined effects of collisional excitation and spontaneous radiative decay.

\noindent 5.  The observed line ratios do not place strong constraints on the gas temperature, and 
thus the inferred physical conditions can plausibly arise behind either dissociative or non-dissociative shocks.

\noindent 6. The collisional excitation model predicts that the 
22~GHz population inversion will be quenched at higher densities than that of the 621~GHz transition, providing a natural explanation for the observational fact that 22~GHz maser emission appears to be a necessary but insufficient condition for 621~GHz maser emission.

\acknowledgments

HIFI has been designed and built by a consortium of institutes and university departments from across
Europe, Canada and the United States under the leadership of SRON Netherlands Institute for Space
Research, Groningen, The Netherlands and with major contributions from Germany, France and the US.
Consortium members are: Canada: CSA, U.~Waterloo; France: CESR, LAB, LERMA, IRAM; Germany:
KOSMA, MPIfR, MPS; Ireland, NUI Maynooth; Italy: ASI, IFSI-INAF, Osservatorio Astrofisico di Arcetri-
INAF; Netherlands: SRON, TUD; Poland: CAMK, CBK; Spain: Observatorio Astron\'omico Nacional (IGN),
Centro de Astrobiolog\'a (CSIC-INTA). Sweden: Chalmers University of Technology - MC2, RSS \& GARD;
Onsala Space Observatory; Swedish National Space Board, Stockholm University - Stockholm Observatory;
Switzerland: ETH Zurich, FHNW; USA: Caltech, JPL, NHSC.

Support for this work was provided by NASA through an award issued by JPL/Caltech.
{This study was partly based on observations with the 100-m telescope of the 
MPIfR (Max-Planck-Institut f\"ur Radioastronomie) at Effelsberg.}  We are grateful to M.~Gerin for useful comments on the manuscript.  We thank S.\ Jones and collaborators for providing us with a draft of their results prior to publication.

\vskip 0.2 true in
\centerline{\bf Appendix: VLA observations of 22 GHz H$_2$O masers }
 
We have downloaded, from the NRAO archive, a set of 22 GHz H$_2$O maser data targeting Orion-KL.
The observations were performed on 1996 March 15 with the VLA in its C configuration, as part of a project (AC443) that targeted seven sources.  The total observing time for
that project  was 3 hours, of which 16 minutes
was devoted to on-source observations of Orion-KL.  The extragalactic radio source 0605-085 was used as a phase calibrator.  In the absence of any non-variable radio source that could be used for absolute flux calibration, we established the flux density scale by assuming a flux density for 0605-085 of 1.7 Jy, the value measured in 1997 December by (Kovalev et al.\ 1999). The spectral-line observations -- carried out 
with the 1.3 cm receivers -- employed a single intermediate frequency band (IF)
mode with total of 127 channels, each of width 49 kHz, corresponding to 0.7 km
s$^{-1}$.  The central channel was set to a local standard of rest (LSR)
velocity of 10 km s$^{-1}$ for the 22~GHz transition, providing a total LSR velocity coverage from $\sim -32$ to +52~km/s.

The data were reduced with Astronomical Image Processing System (AIPS) software
package.  After a standard flux and phase calibration, we used the strongest
maser emission feature - which has a flux of 1500 Jy and a v$_{\rm LSR}$ of --4.6 km s$^{-1}$ -
for self calibration, and then applied the solutions to all the other channels. The
FWHM primary beam of individual VLA antennas is $\approx$ 2$^\prime$. We imaged
and cleaned the data by applying the natural weighting of the UV data, which resulted in a
synthesized beam of size
1\rlap{$^{\prime\prime}$}.3~$\times$~1\rlap{$^{\prime\prime}$}.0 (FWHM) at position angle
--23$^\circ$. We produced a 2048~$\times$2048 maser image cube, with a cell size
of 0\rlap{$^{\prime\prime}$}.1~$\times$~0\rlap{$^{\prime\prime}$}.1.  We then
used the AIPS task SAD to detect maser emissions with peak intensity higher
than 5 times the r.m.s.\ noise level.  For line channels in the Orion-KL
region that are not limited by dynamic range, the r.m.s. noise level was $\approx$ 0.12 Jy. 

In Table 3, we list the parameters obtained for 339 maser emission spots, all selected
to satisfy the following criteria: flux $\geq$ 1 Jy, signal noise ratio (SNR) $\geq$ 10, and
dynamic range $\geq$ 0.01. (In other words, for channels with peak flux greater than 100 Jy, we
used a flux cutoff of 0.01 $\times$ the peak flux). Figure 11
shows distribution of maser spots, with the color and symbol size representing LSR velocity and peak flux respectively.  For comparision with the 621 GHz maser feature 
observed toward Orion-KL,
we show in Figure 12 the 22 GHz maser
emission integrated over the 10~--~13~km s$^{-1}$ velocity range. The letters A, B, C, D, E and
F mark subregions from which the spectra presented in Figure 13 were extracted.
These spectra were obtained at the peak of each subregion, after
Hanning smoothing the data with a kernel of FWHM $1^{\prime\prime}$.

{}

\begin{deluxetable}{llll}

\tablewidth{0pt}
\tabletypesize{\scriptsize}
\tablecaption{Summary of observations} 
\tablehead{Source & Orion-KL & Orion S & W49N \\}
\startdata
R.A. (J2000) 	
& \phantom{--}5h 35m 14.31s $^a$
& \phantom{--}5h 35m 13.44s 
& \phantom{--}19h 10m 13.20s \\
Declination (J2000) 
& --5d 22$^\prime$ 23.2$^{\prime\prime}$ $^a$
& --5d 24$^\prime$  8.1$^{\prime\prime}$
& \phantom{--}9d 06$^\prime$ 12.0$^{\prime\prime}$ \\

\underbar{{\it Herschel} observations (621 GHz transition)}
\\
Date  			& 2011-03-11 	& 2010-08-31 	& 2010-04-11 (1st epoch)\\
      			&            	&            	& 2011-10-08 (2nd epoch)\\
      			&            	&            	& 2012-04-14 (3rd epoch)\\
Total duration (s)  	& 3964 		& 7482 		& 226 $^b$ (1st epoch)\\
  			&        	& 	  	& 158 $^b$ (2nd epoch) \\
  			&        	& 	  	& 282 $^b$ (3rd epoch) \\
\re{R.M.S.}\ noise (mK)	& 65 $^c$ 	& 30 $^d$	&  16 $^d$ (1st epoch)\\
			&               &               &  16 $^d$ (2nd epoch)\\
			&               &               &  12 $^d$ (3rd epoch)\\
HIFI Mode 		& OTF mapping 	& Full spectral scan & Pointed \\
HIFI Key program 	& HEXOS 	& HEXOS 	& PRISMAS/DDT \\
Herschel AORs \#s 	& 1342215918--19& 1342204001	& 1342194523--25 (1st epoch)\\ 
			&		&		& 1342230378--80 (2nd epoch)\\
			& 		& 		& 1342244402--04 (3rd epoch)\\
\\
\underbar{Effelsberg observations (22 GHz transition)}
\\
Date  			& 2011-03-21 	& 2012-9-28  	& 2012-10-01 \\
Receiver		& Prime focus	& Secondary focus &	Prime focus \\
Observing mode	        & Raster map	& On-Off	& On-Off \\
Gain ($T_A/S$)		& 0.9 K/Jy	& 0.9 K/Jy	& 0.9 K/Jy \\	
System temperature (K)$^e$  &	56 	& 97		& 72 \\

\enddata
\tablenotetext{a}{Position of HIFI map center}
\tablenotetext{b}{Divided into three observations with different LO settings (see text)}
\tablenotetext{c}{For a individual spectrum, with a 1.1 MHz channel spacing, in a single map and for a single polarization}
\tablenotetext{d}{For an average spectrum, with a 1.1 MHz channel width, obtained by averaging over all AORs and both polarizations}
\tablenotetext{e}{Average over two polarizations}
\end{deluxetable}

\begin{deluxetable}{lllllll}

\tablewidth{0pt}
\tabletypesize{\scriptsize}
\tablecaption{Maser line fluxes and luminosities} 
\tablehead{Source &$V_{\rm LSR}$ range & 621 GHz flux & 22 GHz flux & $L_p$(621 GHz) & $L_p$(22 GHz) & $L_p$(621 GHz)/\\
                  & (km/s) & ($\rm W\, m^{-2}$) & ($\rm W\,m^{-2}$) & $\rm photon\,s^{-1} $ &  $\rm photon\,s^{-1}$& $L_p$(22 GHz)\\  }
\startdata
Orion-KL & [ --40, 60 ]  			& 4.17(--16)$^a$     & 5.37(--17) 	& 2.08(+45)     & 7.50(+45) 	& 0.28 \\
Orion-KL & [ --40, 60 ]     			& 1.86(--17)$^b$    & 5.37(--17) 	& 9.27(+43)$^b$ & 7.50(+45) 	& 0.012 \\
Orion-KL & [ 10, 13 ] 	 		 	& 1.86(--17)$^b$    & 1.25(--17) 	& 9.27(+43)$^b$ & 1.74(+45) 	& 0.053 \\
Orion-KL & [ 10, 13 ] 				& 1.86(--17)$^b$    & 2.85(--18)$^c$ 	& 9.27(+43)$^b$ & 3.96(+44)$^c$ & 0.23 \\
Orion S 
& [ --30, 30 ]  & 1.67(--17)&  2.07(--18)&  8.33(+43)&  2.88(+44)&  0.29 \\
& [ 4, 7 ] & 8.95(--18)&  2.77(--19)&  4.46(+43)&  3.85(+43)&  1.16 \\
W49N 
& [ --80, 100 ] & 5.22(--15)&  2.09(--16)&  1.97(+49)&  2.21(+49)&  0.89 \\
& [ --10, --5 ] & 1.54(--16)&  4.52(--17)&  5.84(+47)&  4.77(+48)&  0.12 \\
& [ --5,0 ] & 1.79(--16)&  1.43(--17)&  6.75(+47)&  1.51(+48)&  0.45 \\
& [ 0, 5 ] & 2.11(--16)&  1.50(--17)&  7.99(+47)&  1.58(+48)&  0.50 \\
& [ 5, 10 ] & 2.80(--16)&  2.18(--17)&  1.06(+48)&  2.30(+48)&  0.46 \\
& [ 10, 15 ] & 2.49(--16)&  2.36(--17)&  9.41(+47)&  2.49(+48)&  0.38 \\
& [ 15, 20 ] & 2.04(--16)&  6.89(--18)&  7.71(+47)&  7.27(+47)&  1.06 \\

\enddata
\tablenotetext{a}{where a(--b) means $\rm a times 10^{-b}$}
\tablenotetext{b}{Narrow component only (after subtraction of broad component)}
\tablenotetext{c}{22 GHz ``residual'' (after subtraction of point source at [$+15.7^{\prime \prime}$,$+13.6^{\prime \prime}$])}
\end{deluxetable}

\begin{deluxetable}{rrrr|rrrr|rrrr|rrrr}

\tablewidth{0pt}
\tabletypesize{\scriptsize}
\tablecaption{22 GHz Masers in Orion-KL\label{tbl-22GHz}} 
\tablehead{$v_{\rm LSR}$ & $\bigtriangleup x^a$ &$\bigtriangleup y$ & $S^b$ & $v_{\rm LSR}$ & $\bigtriangleup x^a$ &$\bigtriangleup y$ & $S^b$ &
$v_{\rm LSR}$ & $\bigtriangleup x^a$ &$\bigtriangleup y$ & $S^b$ & $v_{\rm LSR}$ & $\bigtriangleup x^a$ &$\bigtriangleup y$ & $S^b$ 
\\ 
 (km/s) & ($^{\prime\prime}$) & ($^{\prime\prime}$) & (Jy) &  (km/s) & ($^{\prime\prime}$) & ($^{\prime\prime}$) & (Jy) &
 (km/s) & ($^{\prime\prime}$) & ($^{\prime\prime}$) & (Jy) &  (km/s) & ($^{\prime\prime}$) & ($^{\prime\prime}$) & (Jy) 
\\}
\startdata
-31.8	&	-2.4	&	-2.8	&	7.9	&	-21.2	&	-0.2	&	-2.4	&	7.1	&	-12.6	&	3.5	&	3.1	&	10.7	&	-6.6	&	-2.5	&	-2.8	&	20.2	\\
-31.8	&	-0.2	&	-2.3	&	4.0	&	-20.5	&	12.2	&	17.0	&	15.2	&	-11.9	&	-2.5	&	-2.8	&	14.9	&	-5.9	&	3.5	&	3.2	&	163.4	\\
-31.1	&	12.2	&	17.0	&	13.2	&	-19.9	&	-2.5	&	-2.8	&	10.9	&	-11.9	&	-0.2	&	-2.4	&	13.2	&	-5.9	&	6.4	&	11.1	&	152.9	\\
-30.5	&	-2.4	&	-2.8	&	8.1	&	-19.9	&	-0.2	&	-2.4	&	7.7	&	-11.2	&	12.2	&	17.0	&	18.2	&	-5.9	&	12.2	&	17.0	&	21.5	\\
-30.5	&	-0.2	&	-2.4	&	4.3	&	-19.2	&	12.2	&	17.0	&	15.5	&	-11.2	&	3.5	&	3.1	&	13.4	&	-5.3	&	3.5	&	3.2	&	333.6	\\
-29.8	&	12.2	&	17.0	&	13.4	&	-19.2	&	3.5	&	3.1	&	4.3	&	-10.6	&	-2.5	&	-2.8	&	15.9	&	-5.3	&	6.4	&	11.1	&	101.3	\\
-29.1	&	-2.4	&	-2.8	&	8.4	&	-18.5	&	-2.5	&	-2.8	&	11.4	&	-10.6	&	-0.2	&	-2.4	&	14.9	&	-5.3	&	-0.2	&	-2.4	&	27.8	\\
-29.1	&	-0.2	&	-2.4	&	4.6	&	-18.5	&	-0.2	&	-2.4	&	8.3	&	-9.9	&	12.2	&	17.0	&	18.9	&	-5.3	&	-2.5	&	-2.7	&	22.6	\\
-28.5	&	12.2	&	17.0	&	13.6	&	-17.9	&	12.2	&	17.0	&	15.9	&	-9.9	&	3.5	&	3.1	&	17.5	&	-4.6	&	3.5	&	3.1	&	1532.0	\\
-27.8	&	-2.4	&	-2.8	&	8.7	&	-17.9	&	3.5	&	3.1	&	5.1	&	-9.9	&	-7.5	&	-2.7	&	8.4	&	-3.9	&	3.5	&	3.1	&	1379.0	\\
-27.8	&	-0.2	&	-2.4	&	5.0	&	-17.2	&	-2.5	&	-2.8	&	12.0	&	-9.2	&	-2.5	&	-2.8	&	17.2	&	-3.3	&	3.5	&	3.2	&	153.3	\\
-27.2	&	12.2	&	17.0	&	13.8	&	-17.2	&	-0.2	&	-2.4	&	9.1	&	-9.2	&	-0.2	&	-2.4	&	16.9	&	-3.3	&	12.2	&	17.0	&	23.8	\\
-26.5	&	-2.4	&	-2.8	&	9.0	&	-16.5	&	12.2	&	17.0	&	16.2	&	-9.2	&	-7.6	&	-2.7	&	6.6	&	-2.6	&	3.4	&	3.0	&	148.6	\\
-26.5	&	-0.2	&	-2.4	&	5.3	&	-16.5	&	3.5	&	3.1	&	6.1	&	-8.6	&	3.6	&	3.2	&	35.5	&	-2.6	&	-0.2	&	-2.4	&	44.5	\\
-25.8	&	12.2	&	17.0	&	14.0	&	-15.9	&	-2.5	&	-2.8	&	12.6	&	-8.6	&	12.2	&	17.0	&	19.7	&	-2.6	&	-2.5	&	-2.8	&	28.2	\\
-25.2	&	-2.4	&	-2.8	&	9.3	&	-15.9	&	-0.2	&	-2.4	&	9.9	&	-8.6	&	-7.6	&	-2.7	&	11.8	&	-2.6	&	6.4	&	11.1	&	5.6	\\
-25.2	&	-0.2	&	-2.4	&	5.7	&	-15.2	&	12.2	&	17.0	&	16.7	&	-7.9	&	-7.6	&	-2.7	&	34.4	&	-1.9	&	12.2	&	17.0	&	26.3	\\
-24.5	&	12.2	&	17.0	&	14.3	&	-15.2	&	3.5	&	3.1	&	7.3	&	-7.9	&	-0.2	&	-2.4	&	19.3	&	-1.3	&	-0.2	&	-2.4	&	63.8	\\
-23.8	&	-2.4	&	-2.8	&	9.7	&	-14.6	&	-2.5	&	-2.8	&	13.2	&	-7.9	&	-2.5	&	-2.8	&	18.6	&	-1.3	&	3.5	&	3.0	&	56.4	\\
-23.8	&	-0.2	&	-2.4	&	6.2	&	-14.6	&	-0.2	&	-2.4	&	10.8	&	-7.3	&	3.5	&	3.2	&	47.0	&	-1.3	&	-2.5	&	-2.7	&	33.3	\\
-23.2	&	12.2	&	17.0	&	14.6	&	-13.9	&	12.2	&	17.0	&	17.2	&	-7.3	&	12.2	&	17.0	&	20.5	&	-1.3	&	-3.2	&	-0.8	&	17.4	\\
-22.5	&	-2.5	&	-2.8	&	10.1	&	-13.9	&	3.5	&	3.1	&	8.8	&	-7.3	&	6.4	&	11.1	&	8.5	&	-0.6	&	12.2	&	17.0	&	28.3	\\
-22.5	&	-0.2	&	-2.4	&	6.6	&	-13.2	&	-2.5	&	-2.8	&	14.0	&	-7.3	&	-7.6	&	-2.7	&	4.2	&	0.0	&	-0.2	&	-2.4	&	113.8	\\
-21.8	&	12.2	&	17.0	&	14.9	&	-13.2	&	-0.2	&	-2.4	&	11.9	&	-6.6	&	3.5	&	3.5	&	29.7	&	0.7	&	-3.1	&	-0.8	&	37.3	\\
-21.2	&	-2.5	&	-2.8	&	10.5	&	-12.6	&	12.2	&	17.0	&	17.7	&	-6.6	&	-0.2	&	-2.4	&	22.8	&	0.7	&	12.2	&	17.0	&	31.2	\\
\tablebreak																															
1.4	&	-0.2	&	-2.4	&	1207.0	&	6.7	&	-3.6	&	-5.8	&	457.9	&	11.3	&	-2.5	&	-2.7	&	57.3	&	14.6	&	-7.7	&	-5.3	&	175.2	\\
1.4	&	-2.6	&	-2.7	&	36.7	&	7.3	&	-2.5	&	-2.8	&	1372.0	&	12.0	&	9.7	&	14.5	&	50.4	&	14.6	&	22.8	&	-6.9	&	98.6	\\
2.0	&	-0.2	&	-2.4	&	1158.0	&	7.3	&	12.2	&	17.0	&	55.7	&	12.0	&	-0.2	&	-6.6	&	41.9	&	14.6	&	12.2	&	17.0	&	18.6	\\
2.0	&	12.2	&	17.0	&	33.0	&	8.0	&	-3.6	&	-5.8	&	890.1	&	12.0	&	-7.6	&	-5.4	&	14.0	&	14.6	&	-5.8	&	-9.7	&	8.6	\\
2.7	&	-0.2	&	-2.2	&	520.2	&	8.0	&	0.6	&	-4.4	&	173.6	&	12.0	&	8.5	&	17.2	&	8.9	&	15.3	&	22.8	&	-6.9	&	272.4	\\
2.7	&	-2.7	&	-2.7	&	190.6	&	8.0	&	6.2	&	6.0	&	51.9	&	12.0	&	22.8	&	-6.9	&	6.6	&	15.3	&	-7.7	&	-5.2	&	157.9	\\
2.7	&	12.0	&	-5.2	&	131.1	&	8.7	&	-3.5	&	-5.9	&	1163.0	&	12.7	&	-11.8	&	34.0	&	307.3	&	15.3	&	3.4	&	3.6	&	44.8	\\
3.4	&	-0.2	&	-2.3	&	626.0	&	8.7	&	12.2	&	17.0	&	151.9	&	12.7	&	9.7	&	14.5	&	67.8	&	15.3	&	-3.7	&	-5.6	&	34.5	\\
3.4	&	12.0	&	-5.2	&	64.7	&	8.7	&	-2.5	&	-2.7	&	151.3	&	12.7	&	12.4	&	16.9	&	39.2	&	15.3	&	-2.5	&	-2.8	&	24.4	\\
3.4	&	12.2	&	17.0	&	40.3	&	8.7	&	6.2	&	6.0	&	112.2	&	12.7	&	-2.5	&	-2.8	&	37.5	&	15.3	&	18.1	&	1.2	&	16.1	\\
4.0	&	-3.6	&	-5.8	&	607.0	&	8.7	&	0.5	&	-4.4	&	71.5	&	12.7	&	-0.2	&	-6.6	&	19.4	&	15.3	&	-0.1	&	-2.4	&	16.0	\\
4.0	&	-0.5	&	-2.0	&	110.9	&	8.7	&	20.6	&	-3.9	&	62.3	&	12.7	&	-0.2	&	-2.4	&	16.8	&	16.0	&	-7.6	&	-5.4	&	295.0	\\
4.0	&	3.4	&	1.5	&	67.7	&	9.3	&	-4.1	&	-5.1	&	683.6	&	13.3	&	8.5	&	17.2	&	212.5	&	16.0	&	22.8	&	-6.8	&	28.9	\\
4.0	&	-2.4	&	-2.8	&	63.8	&	9.3	&	20.7	&	-4.0	&	215.9	&	13.3	&	-11.9	&	34.1	&	121.4	&	16.0	&	12.2	&	17.0	&	15.2	\\
4.0	&	3.6	&	3.2	&	21.9	&	9.3	&	2.9	&	3.3	&	81.1	&	13.3	&	-7.9	&	-5.3	&	77.3	&	16.6	&	-7.6	&	-5.4	&	110.6	\\
4.7	&	-0.3	&	-2.2	&	555.0	&	9.3	&	6.2	&	6.0	&	53.7	&	13.3	&	12.3	&	16.9	&	29.1	&	16.6	&	-3.7	&	-5.6	&	27.2	\\
4.7	&	-3.5	&	-5.9	&	439.5	&	9.3	&	-0.2	&	-6.6	&	34.6	&	13.3	&	22.8	&	-6.9	&	12.8	&	16.6	&	-2.5	&	-2.8	&	21.7	\\
4.7	&	12.2	&	17.0	&	49.6	&	9.3	&	-11.8	&	34.0	&	30.0	&	13.3	&	17.2	&	13.9	&	12.3	&	16.6	&	-0.1	&	-2.4	&	14.2	\\
5.4	&	-0.4	&	-2.1	&	218.6	&	10.0	&	-4.0	&	-5.1	&	1331.0	&	13.3	&	-0.1	&	-6.6	&	11.4	&	16.6	&	22.8	&	-6.8	&	7.4	\\
5.4	&	-2.4	&	-2.8	&	135.0	&	10.0	&	12.2	&	17.0	&	548.2	&	14.0	&	2.9	&	3.1	&	324.7	&	17.3	&	-7.7	&	-5.4	&	49.2	\\
5.4	&	3.5	&	3.1	&	17.9	&	10.0	&	-36.6	&	-49.2	&	101.8	&	14.0	&	8.5	&	17.2	&	177.0	&	17.3	&	12.2	&	16.9	&	11.6	\\
6.0	&	12.2	&	17.0	&	64.1	&	10.7	&	12.2	&	17.0	&	1573.0	&	14.0	&	-7.9	&	-5.2	&	47.2	&	18.0	&	-2.5	&	-2.8	&	19.0	\\
6.0	&	-0.2	&	-2.3	&	54.4	&	11.3	&	12.2	&	17.0	&	736.1	&	14.0	&	-2.5	&	-2.8	&	30.8	&	18.0	&	5.8	&	22.7	&	12.8	\\
6.0	&	20.9	&	8.8	&	34.4	&	11.3	&	-11.8	&	34.1	&	593.2	&	14.0	&	17.2	&	13.8	&	21.7	&	18.6	&	-7.7	&	-5.4	&	170.6	\\
6.7	&	-2.5	&	-2.8	&	767.3	&	11.3	&	-0.2	&	-6.6	&	67.1	&	14.0	&	-0.2	&	-2.4	&	18.0	&	18.6	&	0.7	&	-3.9	&	41.3	\\
\tablebreak																															
18.6	&	12.2	&	16.9	&	8.4	&	22.6	&	-14.0	&	13.3	&	10.7	&	27.3	&	-7.7	&	-5.4	&	7.0	&	34.6	&	3.5	&	3.0	&	6.3	\\
19.3	&	-7.7	&	-5.4	&	141.6	&	22.6	&	3.5	&	3.0	&	8.3	&	27.3	&	0.8	&	-4.5	&	3.9	&	35.2	&	-0.2	&	-2.4	&	6.0	\\
19.3	&	-3.0	&	-5.8	&	94.4	&	22.6	&	-8.4	&	14.7	&	5.9	&	27.9	&	3.5	&	3.0	&	7.2	&	35.2	&	-2.6	&	-2.8	&	5.2	\\
19.3	&	0.6	&	-3.9	&	73.6	&	22.6	&	12.3	&	16.9	&	3.7	&	28.6	&	-2.5	&	-2.7	&	9.0	&	35.2	&	-7.7	&	-5.4	&	4.1	\\
19.3	&	-2.5	&	-2.8	&	17.3	&	23.3	&	-3.6	&	-5.7	&	13.0	&	28.6	&	-3.6	&	-5.8	&	8.0	&	35.2	&	12.1	&	17.0	&	3.1	\\
19.3	&	-0.2	&	-2.3	&	10.9	&	23.3	&	-2.5	&	-2.7	&	11.8	&	28.6	&	-0.2	&	-2.4	&	7.6	&	35.9	&	-10.6	&	11.6	&	35.0	\\
20.0	&	-2.8	&	-5.8	&	136.2	&	23.3	&	-7.7	&	-5.4	&	11.7	&	28.6	&	-7.7	&	-5.4	&	6.2	&	35.9	&	3.5	&	3.0	&	6.2	\\
20.0	&	0.6	&	-3.9	&	36.1	&	23.3	&	-0.2	&	-2.4	&	9.6	&	29.2	&	3.5	&	3.0	&	7.0	&	36.5	&	-10.6	&	11.6	&	203.0	\\
20.0	&	3.5	&	3.0	&	9.1	&	23.9	&	3.5	&	3.0	&	7.9	&	29.9	&	-2.5	&	-2.7	&	8.6	&	36.5	&	-2.6	&	-2.8	&	6.2	\\
20.0	&	12.2	&	16.9	&	7.1	&	23.9	&	12.3	&	16.9	&	2.5	&	29.9	&	-0.2	&	-2.4	&	7.3	&	36.5	&	-0.1	&	-2.3	&	5.3	\\
20.6	&	-3.0	&	-5.8	&	63.5	&	24.6	&	-3.6	&	-5.8	&	11.3	&	29.9	&	-3.6	&	-5.9	&	7.1	&	37.2	&	-10.6	&	11.6	&	122.8	\\
20.6	&	-7.7	&	-5.4	&	27.4	&	24.6	&	-2.5	&	-2.8	&	10.4	&	29.9	&	-7.7	&	-5.4	&	5.6	&	37.2	&	3.5	&	3.0	&	6.2	\\
20.6	&	-2.5	&	-2.7	&	15.8	&	24.6	&	-7.7	&	-5.4	&	9.4	&	30.6	&	3.5	&	3.0	&	6.8	&	37.2	&	-10.5	&	9.3	&	3.3	\\
20.6	&	-0.2	&	-2.4	&	10.9	&	24.6	&	-0.2	&	-2.4	&	9.0	&	31.2	&	-2.5	&	-2.7	&	10.8	&	37.9	&	-0.2	&	-2.4	&	5.5	\\
21.3	&	0.8	&	-4.4	&	11.9	&	24.6	&	0.9	&	-4.6	&	5.2	&	31.2	&	-0.2	&	-2.4	&	6.9	&	37.9	&	-2.6	&	-2.8	&	4.4	\\
21.3	&	3.5	&	3.0	&	8.7	&	25.3	&	3.5	&	3.0	&	7.6	&	31.2	&	-7.7	&	-5.4	&	5.1	&	37.9	&	12.1	&	17.0	&	3.8	\\
21.3	&	12.3	&	16.9	&	5.1	&	25.3	&	12.4	&	16.8	&	1.5	&	31.9	&	3.5	&	3.0	&	6.6	&	37.9	&	-7.7	&	-5.4	&	3.7	\\
21.3	&	-5.2	&	9.5	&	1.8	&	25.9	&	-3.6	&	-5.8	&	10.0	&	32.6	&	-2.6	&	-2.7	&	7.2	&	38.5	&	3.5	&	3.0	&	5.8	\\
21.9	&	-14.1	&	13.3	&	23.0	&	25.9	&	-2.5	&	-2.8	&	9.5	&	32.6	&	-0.2	&	-2.4	&	6.6	&	38.5	&	-10.6	&	11.6	&	3.7	\\
21.9	&	-3.5	&	-5.7	&	17.3	&	25.9	&	-0.2	&	-2.4	&	8.5	&	32.6	&	-7.7	&	-5.4	&	4.7	&	39.2	&	-0.2	&	-2.4	&	5.3	\\
21.9	&	-2.6	&	-2.7	&	16.1	&	25.9	&	-7.7	&	-5.4	&	8.0	&	33.2	&	3.5	&	3.0	&	6.4	&	39.2	&	12.1	&	17.0	&	4.1	\\
21.9	&	-7.7	&	-5.4	&	15.8	&	26.6	&	3.5	&	3.0	&	7.4	&	33.9	&	-0.2	&	-2.4	&	6.3	&	39.2	&	-2.6	&	-2.8	&	3.9	\\
21.9	&	-0.2	&	-2.4	&	10.2	&	27.3	&	-3.6	&	-5.8	&	8.9	&	33.9	&	-2.6	&	-2.8	&	5.7	&	39.2	&	-7.7	&	-5.4	&	3.5	\\
21.9	&	0.8	&	-4.6	&	5.9	&	27.3	&	-2.5	&	-2.8	&	8.6	&	33.9	&	-7.7	&	-5.4	&	4.4	&	39.9	&	3.5	&	3.0	&	5.7	\\
21.9	&	-8.3	&	14.8	&	5.2	&	27.3	&	-0.2	&	-2.4	&	8.1	&	33.9	&	12.1	&	17.0	&	2.7	&	39.9	&	-10.6	&	11.6	&	2.2	\\
\tablebreak																															
40.5	&	-0.2	&	-2.4	&	5.1	&	42.5	&	3.5	&	3.0	&	5.5	&	45.8	&	12.2	&	17.0	&	5.4	&	48.5	&	-0.2	&	-2.4	&	4.0	\\
40.5	&	12.2	&	17.0	&	4.4	&	43.2	&	12.2	&	17.0	&	4.9	&	45.8	&	-0.2	&	-2.4	&	4.4	&	49.1	&	3.5	&	3.0	&	5.0	\\
40.5	&	-2.6	&	-2.8	&	3.6	&	43.2	&	-0.2	&	-2.4	&	4.7	&	45.8	&	-7.7	&	-5.4	&	2.9	&	49.8	&	12.2	&	17.0	&	5.9	\\
40.5	&	-7.7	&	-5.4	&	3.4	&	43.2	&	-7.7	&	-5.4	&	3.2	&	45.8	&	-2.6	&	-2.8	&	2.5	&	49.8	&	-0.2	&	-2.4	&	3.8	\\
41.2	&	3.5	&	3.0	&	5.6	&	43.2	&	-2.6	&	-2.8	&	3.0	&	46.5	&	3.5	&	3.0	&	5.2	&	50.5	&	3.5	&	3.0	&	4.9	\\
41.2	&	-10.6	&	11.6	&	1.5	&	43.8	&	3.5	&	3.0	&	5.4	&	47.2	&	12.2	&	17.0	&	5.6	&	51.1	&	12.2	&	17.0	&	6.1	\\
41.8	&	-0.2	&	-2.4	&	4.9	&	44.5	&	12.2	&	17.0	&	5.1	&	47.2	&	-0.2	&	-2.4	&	4.2	&	51.1	&	-0.2	&	-2.4	&	3.7	\\
41.8	&	12.2	&	17.0	&	4.6	&	44.5	&	-0.2	&	-2.4	&	4.5	&	47.2	&	-7.7	&	-5.4	&	2.9	&	51.1	&	-2.7	&	-2.8	&	1.7	\\
41.8	&	-2.6	&	-2.8	&	3.3	&	44.5	&	-2.6	&	-2.8	&	2.7	&	47.8	&	3.5	&	3.0	&	5.1	&	51.8	&	3.5	&	3.0	&	4.8	\\
41.8	&	-7.7	&	-5.4	&	3.3	&	45.2	&	3.5	&	3.0	&	5.3	&	48.5	&	12.2	&	17.0	&	5.8	\\								
\enddata
\tablenotetext{a}{Position offset relative to 
05h35m14.3s, -05d22$^\prime$'33.7$^{\prime\prime}$ (J2000).}
\tablenotetext{b}{Peak flux}

\end{deluxetable}

\begin{figure}
\includegraphics[width=11 cm, angle=90]{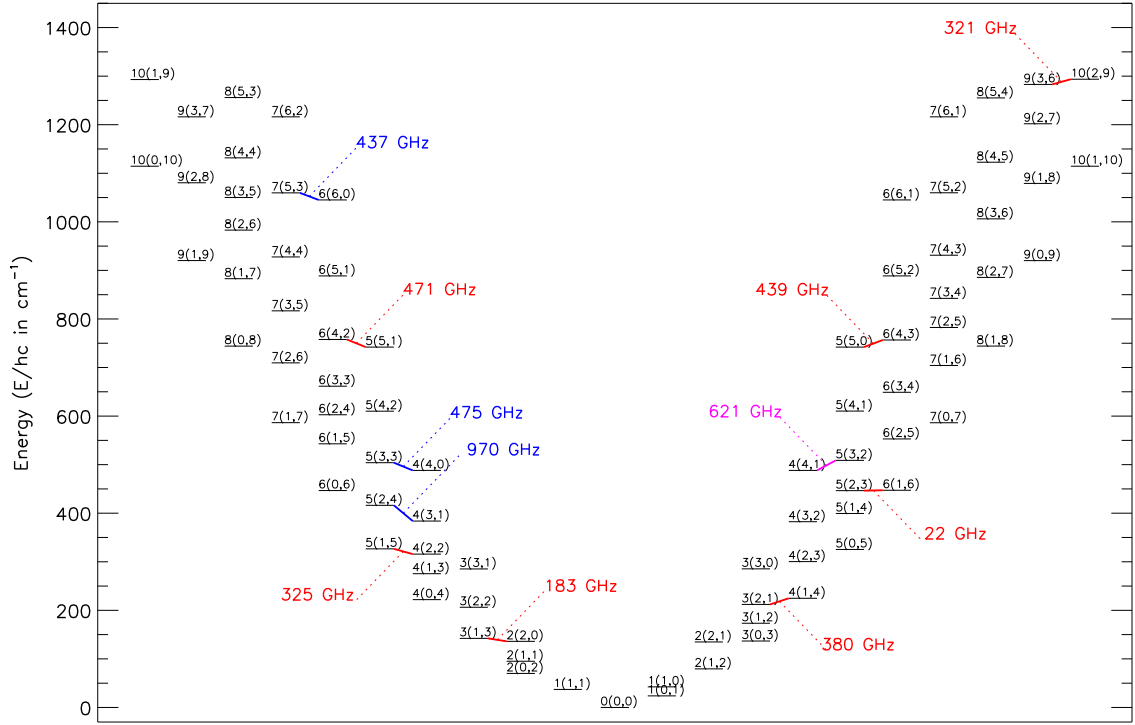}
\caption{Energy level diagram for the ground vibrational state of water.  Red lines: maser transitions observed in both circumstellar and interstellar gas. Blue lines: maser transition observed in circumstellar gas alone. Magenta line: 621 GHz maser transition studied here.}
\end{figure}

\begin{figure}
\includegraphics[width=10 cm]{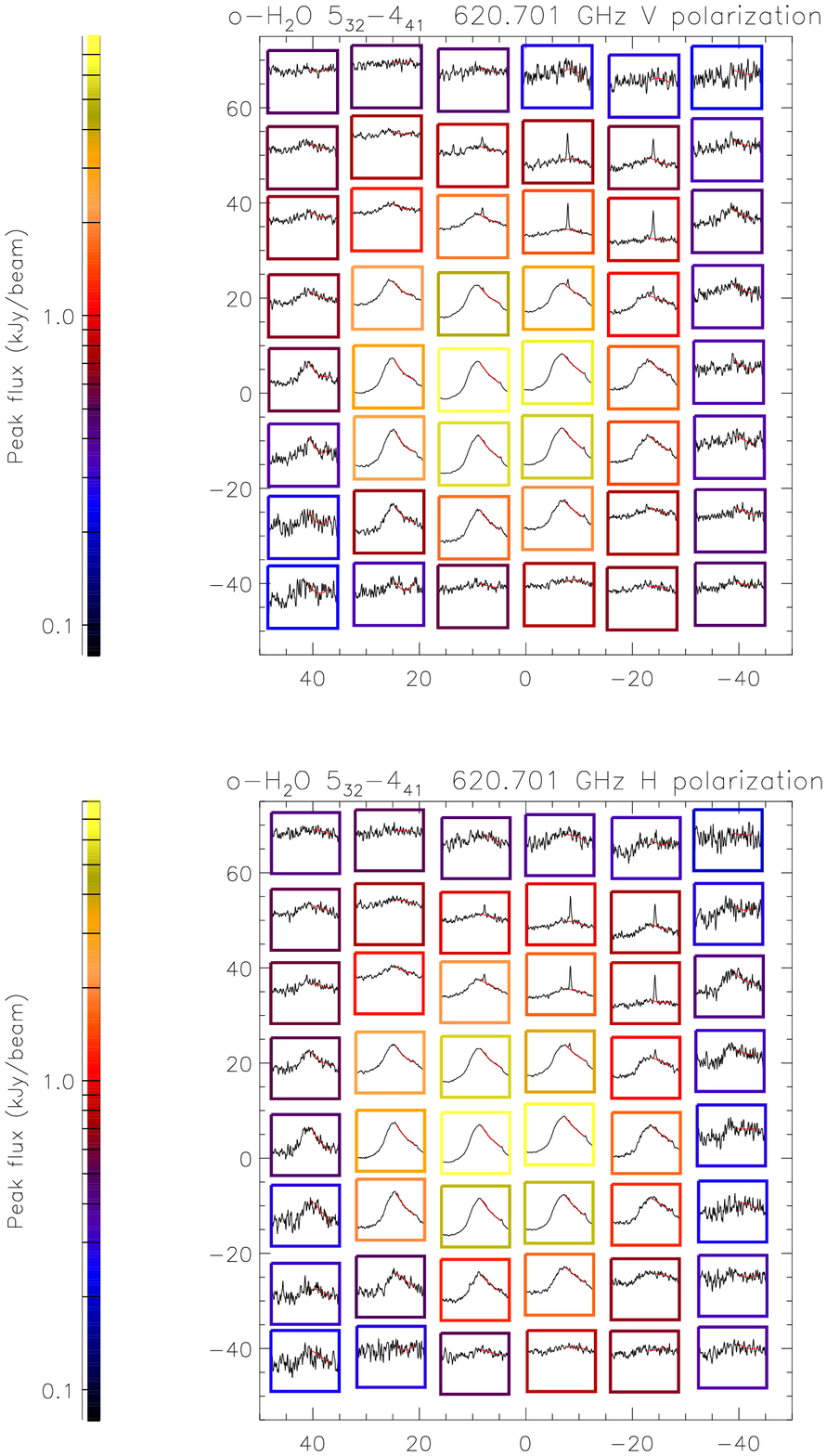}
\caption{621 GHz maser spectra, obtained from mapping observations of Orion-KL.  The spectra are autoscaled, with the color of each border indicating the \re{peak of the vertical scale}.  Two linear polarizations are shown in separate panels.  {Offsets are shown in arcsec, computed relative to Orion-KL at position $\rm 5h\,35m\,14.3s, -5d\,22^\prime\,33.7^{\prime \prime}$ (J2000).}}
\end{figure}

\begin{figure}
\includegraphics[width=10 cm]{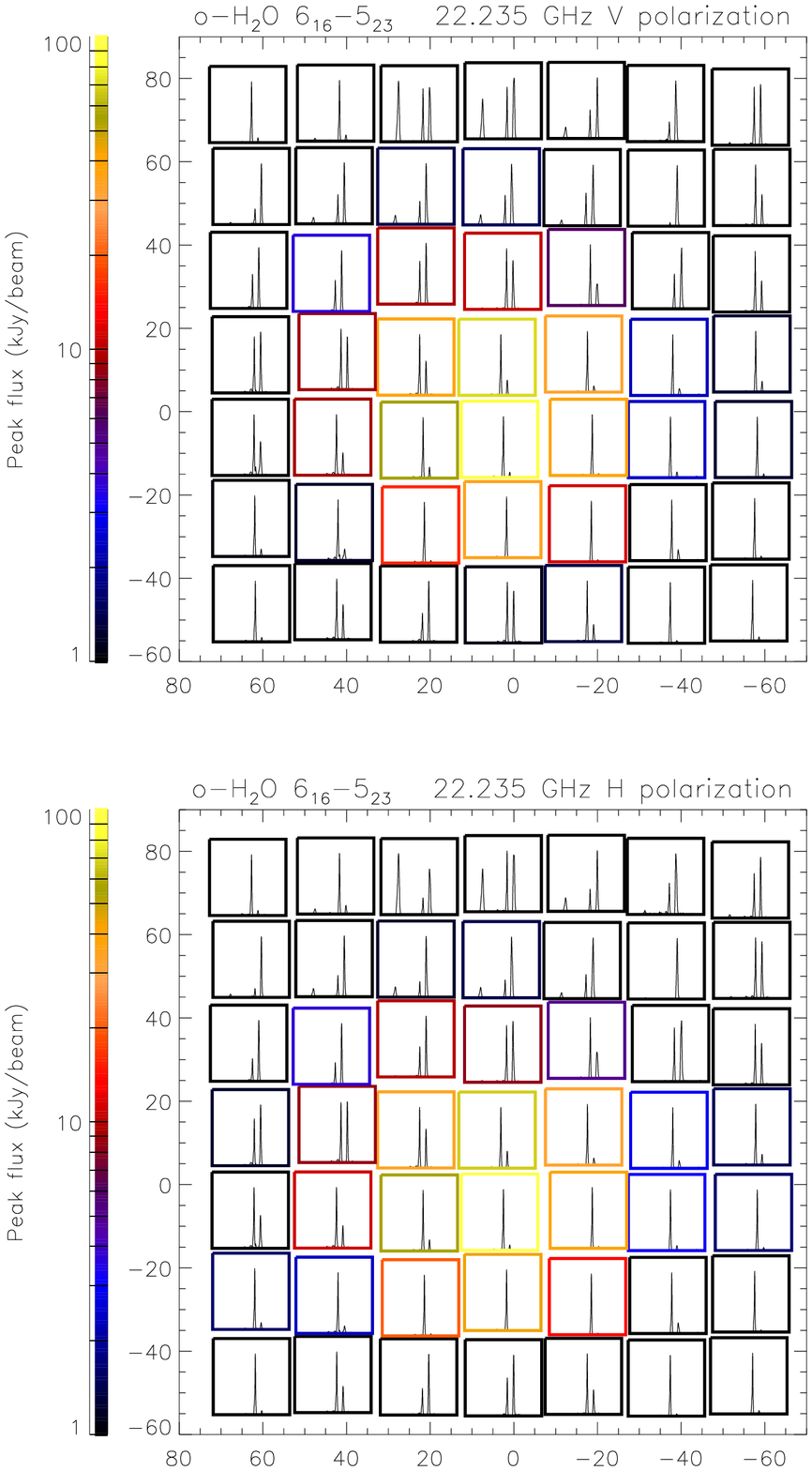}
\caption{22 GHz maser spectra, obtained from mapping observations of Orion-KL.  The spectra are autoscaled, with the color of each border indicating the \re{peak of the vertical scale}.  Two linear polarizations are shown in separate panels.   {Offsets are shown in arcsec, computed relative to Orion-KL at position $\rm 5h\,35m\,14.3s, -5d\,22^\prime\,33.7^{\prime \prime}$ (J2000).}}   
\end{figure}

\begin{figure}
\includegraphics[width=14 cm]{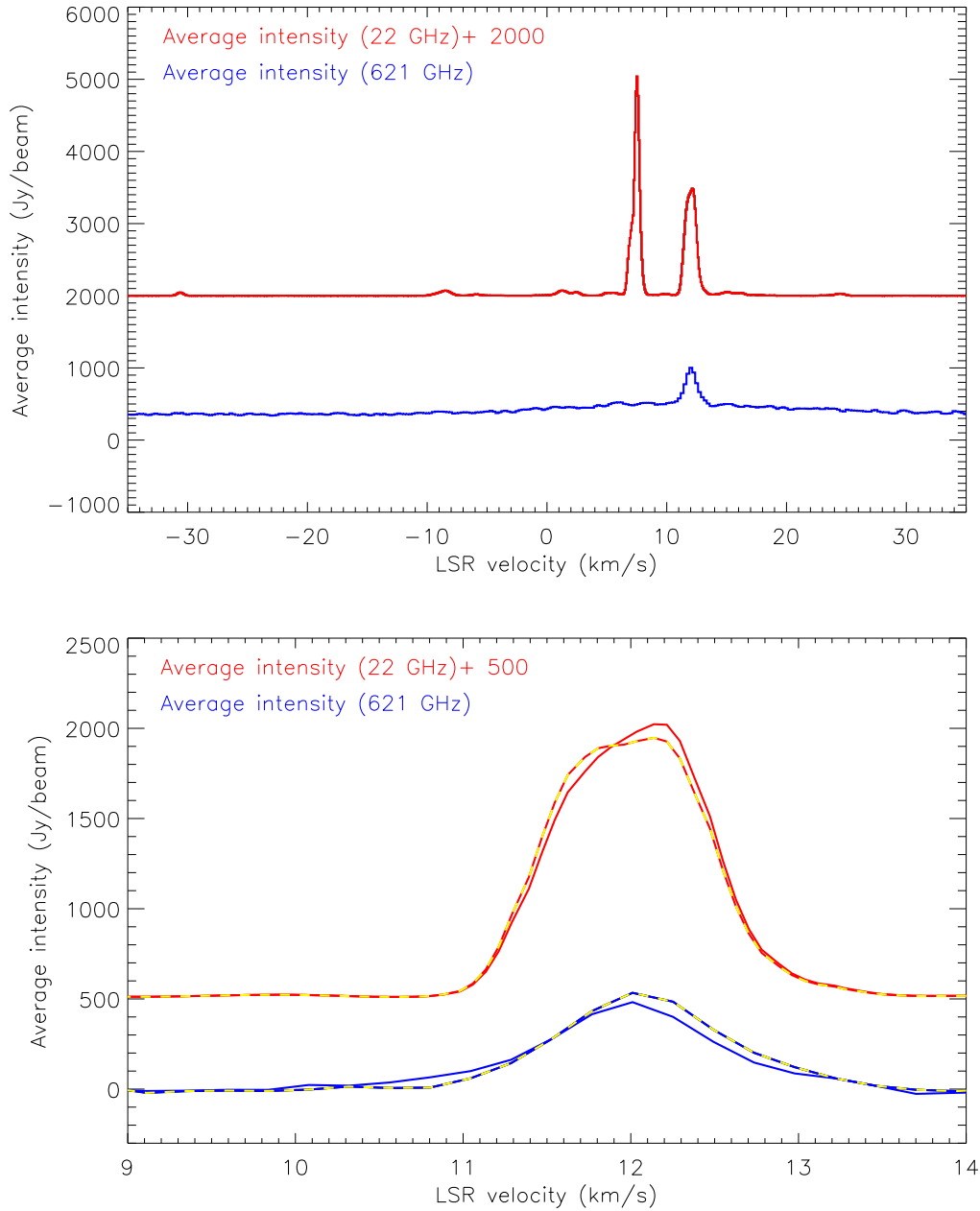}
\caption{Average spectra obtained near offset position [$-15^{\prime \prime}$,$43^{\prime \prime}$](average of four closest positions for the 621 GHz line or 2 closest positions for the 22 GHz line).  
Lower panel: the 9 -- 14 km/s velocity range, after subtraction of a broad pedestal (621 GHz line), and with two linear polarizations shown separately. }
\end{figure}

\begin{figure}
\includegraphics[width=14 cm]{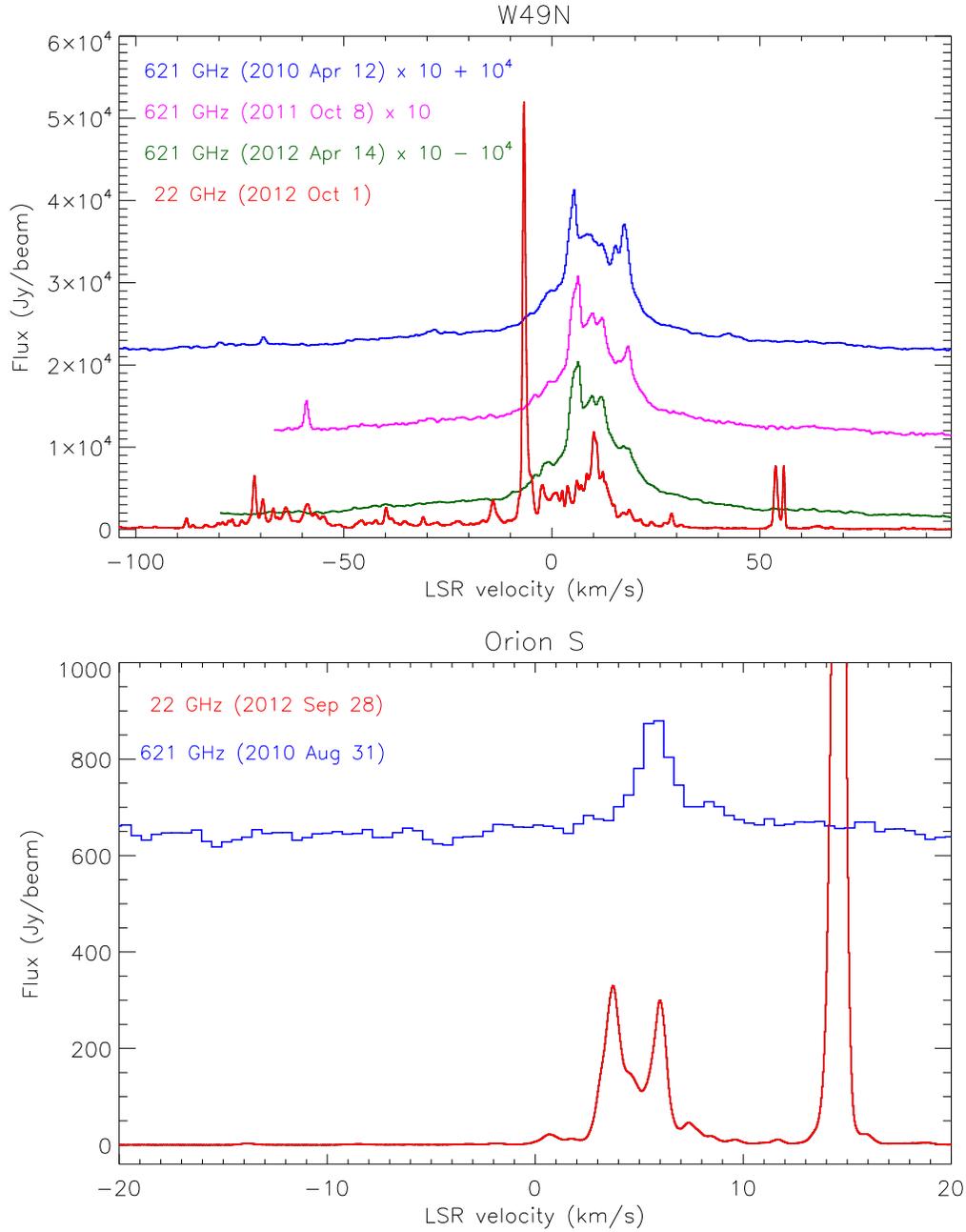}
\caption{22 GHz (red) and 621 GHz (blue, magenta, green) spectra obtained toward W49N and Orion S.  The peak flux for the 22 GHz line toward Orion S was 2900~Jy/beam. }
\end{figure}

\begin{figure}
\includegraphics[width=15 cm]{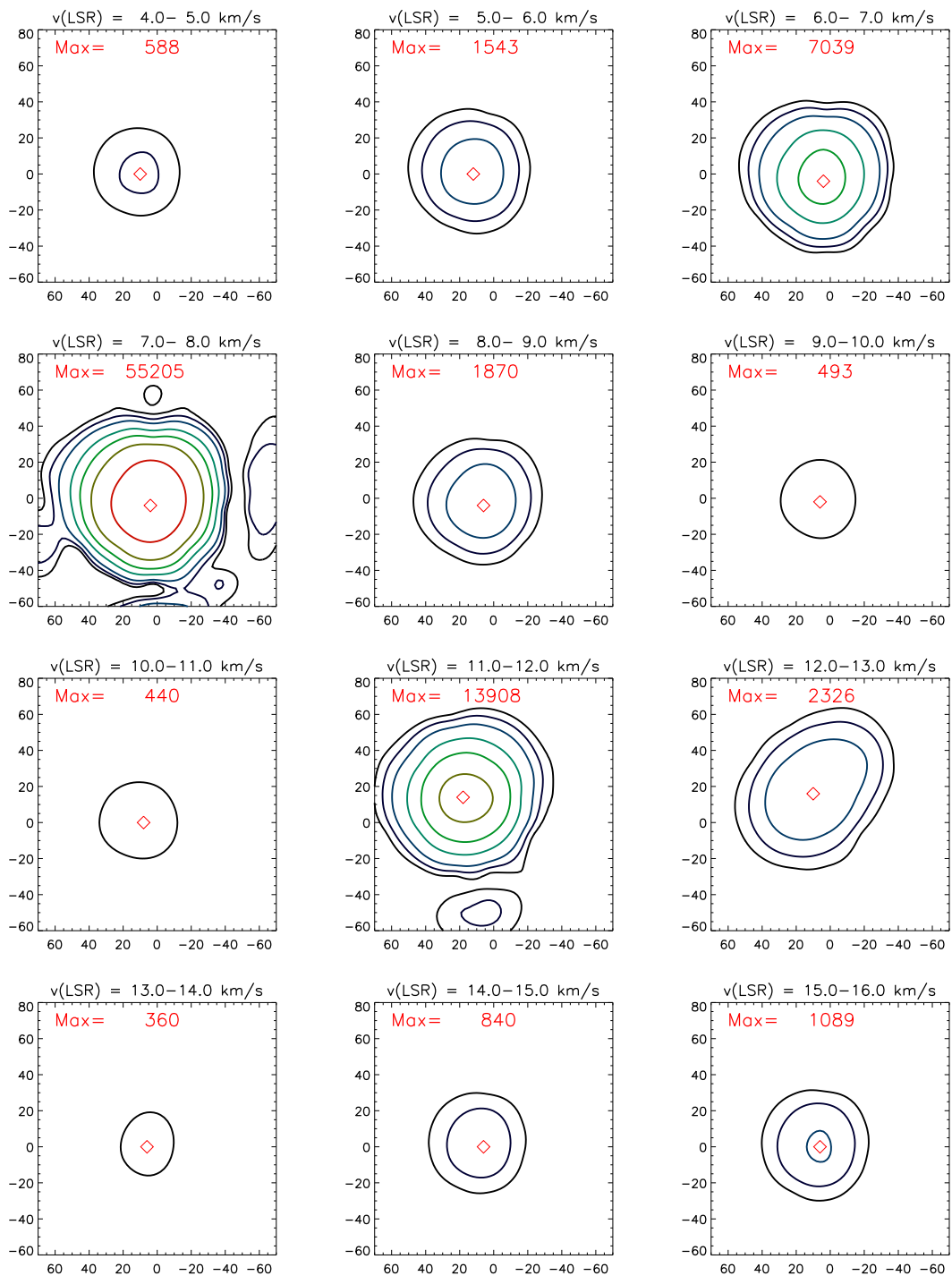}
\caption{Channel maps obtained for the 22 GHz maser transition in Orion-KL.  {Offsets are shown in arcsec, computed relative to Orion-KL at position $\rm 5h\,35m\,14.3s, -5d\,22^\prime\,33.7^{\prime \prime}$ (J2000).  The location of the peak intensity is marked by a red diamond.}}
\end{figure}

\begin{figure}
\includegraphics[width=7.5 cm]{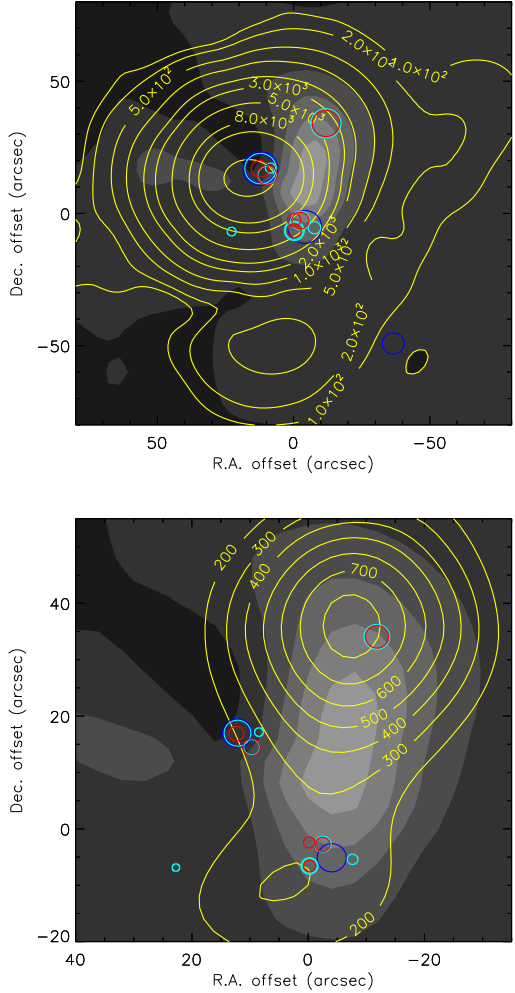}
\caption{Yellow contours: 22 GHz (upper panel) and 621 GHz (lower panel) intensities (Jy km/s per beam), integrated over the 10 -- 13 km/s LSR velocity range, measured toward Orion-KL.  Note that the lower panel covers a smaller region than the upper panel.
Grayscale in both panels: 22 GHz residuals relative to the beam response for a point source at offset position ($+15.7^{\prime \prime}$,$+13.6^{\prime \prime}$). Successive grayscale values are separated by 1000 Jy km/s per beam, with the lowest value corresponding to negative residuals in the interval [--1000 Jy km/s per beam, 0].  \re{Open circles show the locations of 22 GHz maser spots observed in VLA observations performed on 1996 March 15 (see Appendix).
Blue, cyan and red circles indicate LSR velocities in the 10 -- 11, 11 -- 12, and 12 -- 13~km/s ranges, respectively, with the size of each circle indicating the flux on a logarithmic scale.}  Offsets are shown in arcsec, computed relative to Orion-KL at position $\rm 5h\,35m\,14.3s, -5d\,22^\prime\,33.7^{\prime \prime}$ (J2000).}

\end{figure}
\begin{figure}
\includegraphics[width=8.5 cm]{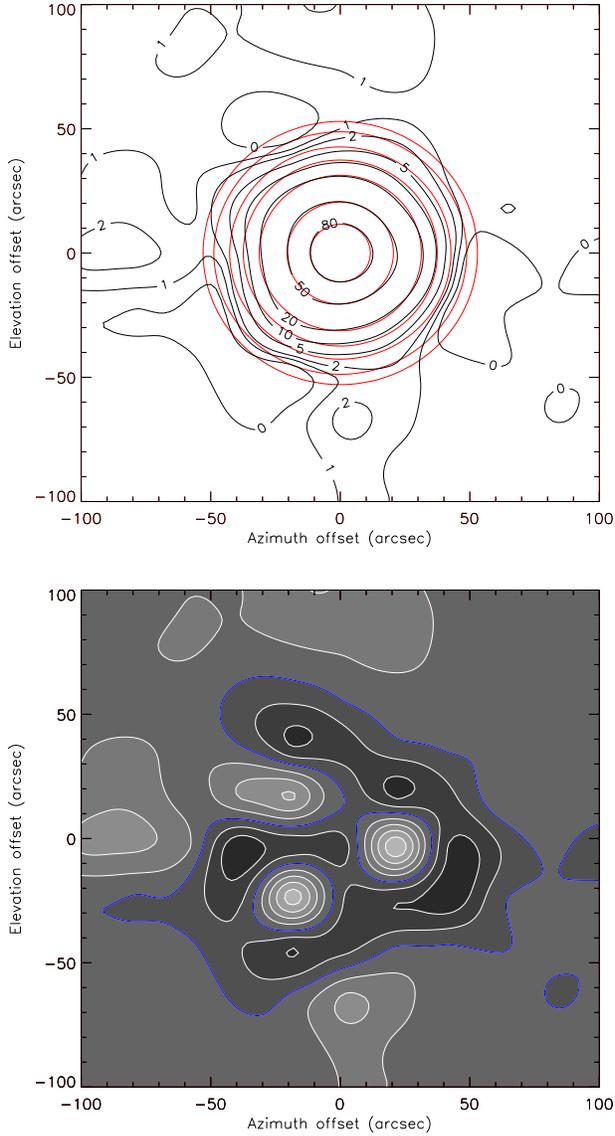}
\caption{Upper panel: observed beam profile for the Effelsberg 100 m telescope at 22 GHz (black contours, showing the response as a percentage of the peak.  The red contours show the best-fit Gaussian, with a HPBW of 41.03$^{\prime\prime}$.
Lower panel: residuals after subtraction of the best-fit Gaussian, with the contours spaced by $1\%$ of the peak beam response and the zero level shown by the blue contour.   
}

\end{figure}

\begin{figure}
\includegraphics[width=12 cm, angle=90]{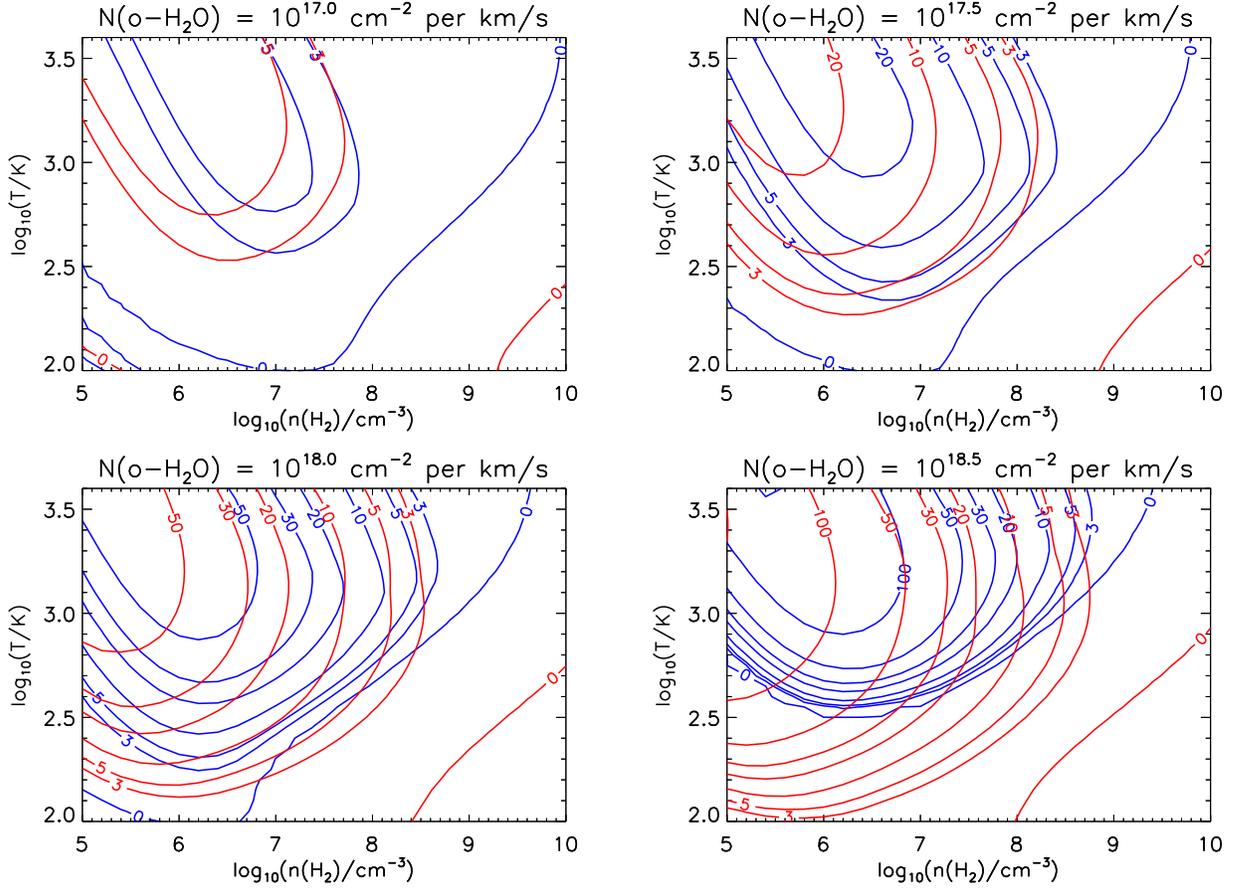} 
\caption{Contours of predicted $-\rm \tau_S$, where $\rm \tau_S$ is the Sobolev optical depth.  These optical depths apply in the direction of the velocity gradient (for which the magnitude of the optical depth is smallest).  Red and blue contours apply to the 22~GHz and 621~GHz maser transitions { respectively}.}
\end{figure}

\begin{figure}
\includegraphics[width=12 cm, angle=90]{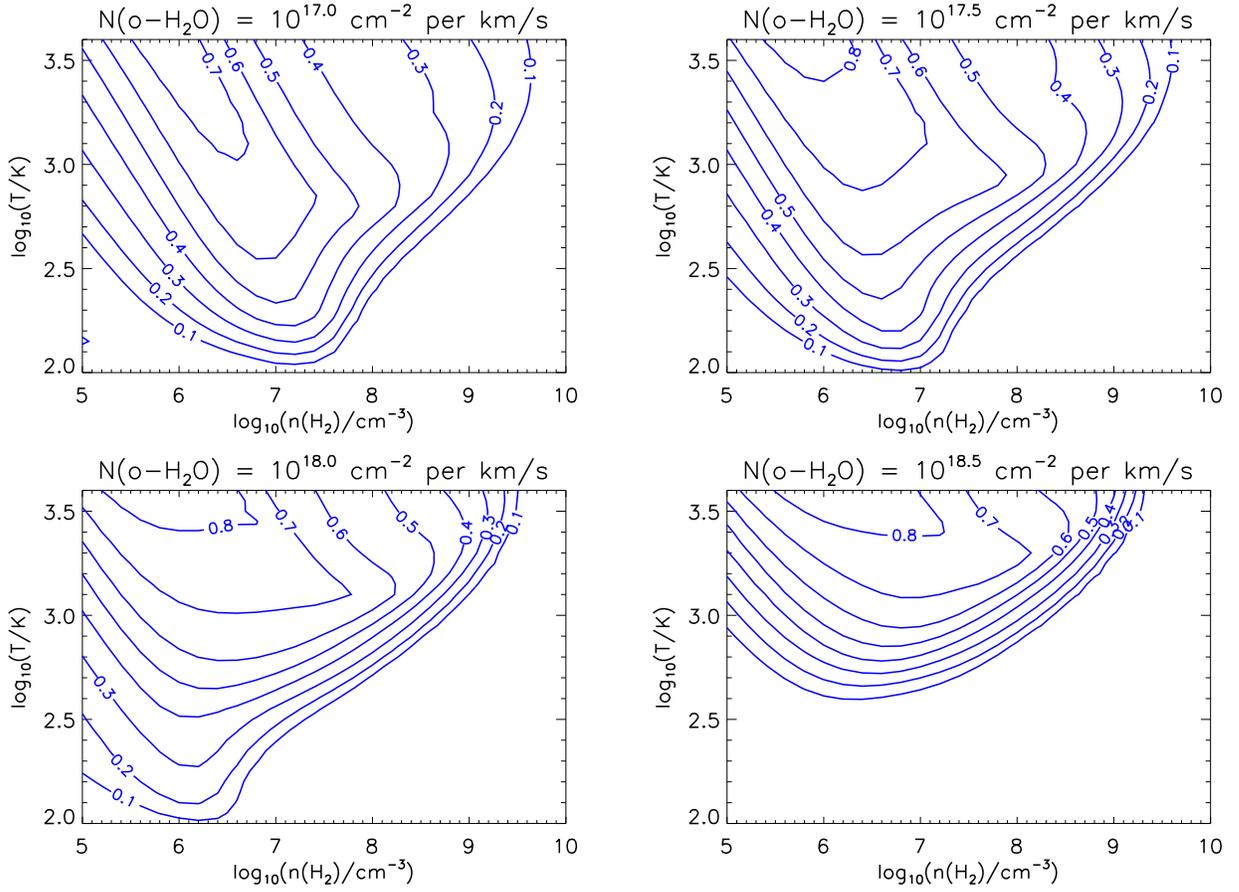}
\caption{Predicted ratio of 621 GHz to 22 GHz maser photon luminosities (see text)}
\end{figure}

\begin{figure} 
\includegraphics[width=11 cm, angle=-90]{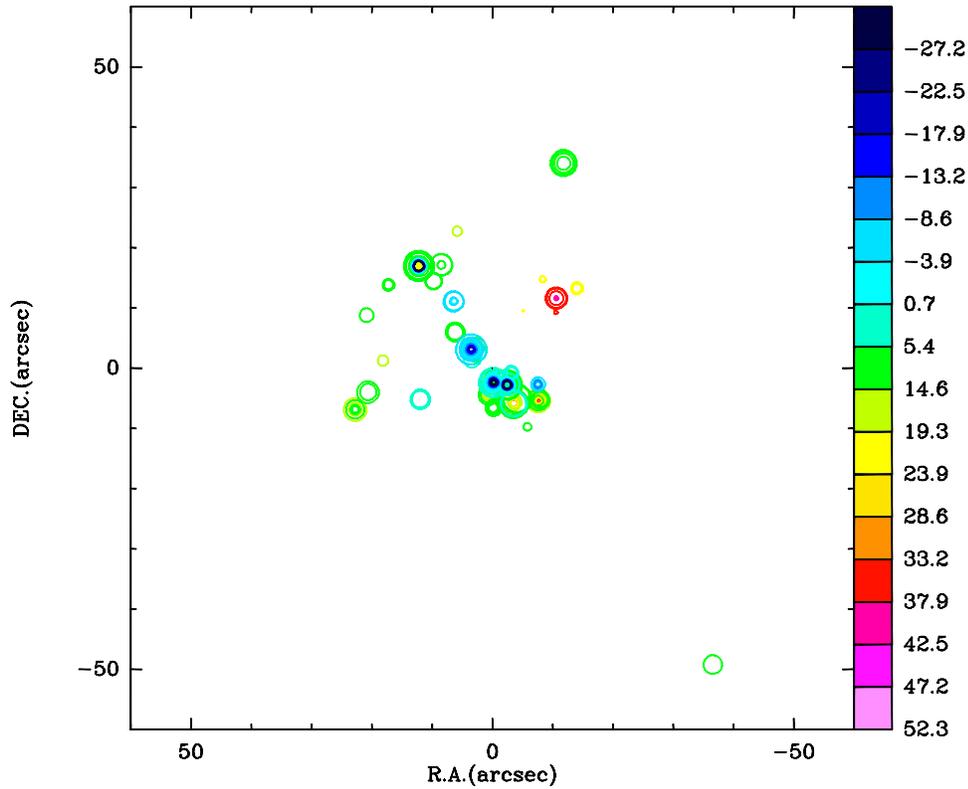}
\caption{22 GHz maser spot distribution in Orion-KL, obtained from archival VLA observations performed on 1996 March 15.  The symbol size increases logarithmically
with the maser flux, and the symbol color denotes the velocity of the maser
spots. Offsets are shown in arcsec, computed relative to Orion-KL at position $\rm 5h\,35m\,14.3s, -5d\,22^\prime\,33.7^{\prime \prime}$ (J2000). \label{fig-spotsmap}} 
\end{figure}

\begin{figure} 
\includegraphics[width=11 cm, angle=-90]{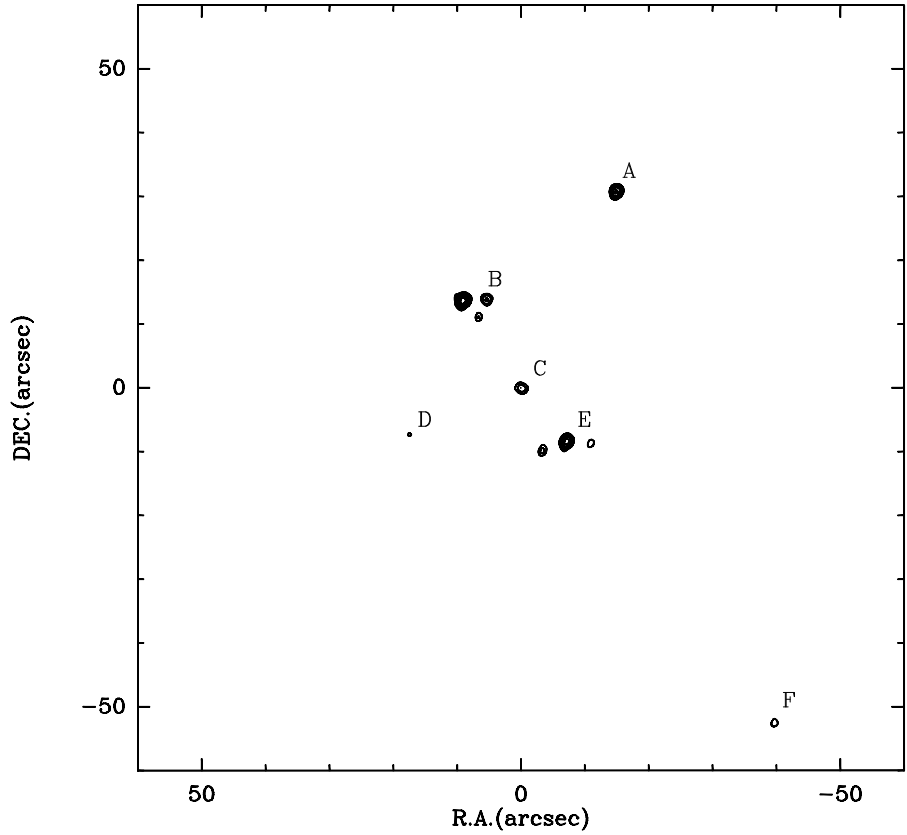}
\caption{Integrated intensity of 22 GHz maser over the velocity range~10~--~13~km
s$^{-1}$, obtained from the 1996 VLA observations, with contour levels 30~$\times$~(1, 2, 4, 8, 16, 32) Jy km
s$^{-1}$ per beam.  Offsets are shown in arcsec, computed relative to Orion-KL at position $\rm 5h\,35m\,14.3s, -5d\,22^\prime\,33.7^{\prime \prime}$ (J2000).\label{fig-10-13kms}} 
\end{figure}

\begin{figure} 
\includegraphics[width=5 cm, angle=0]{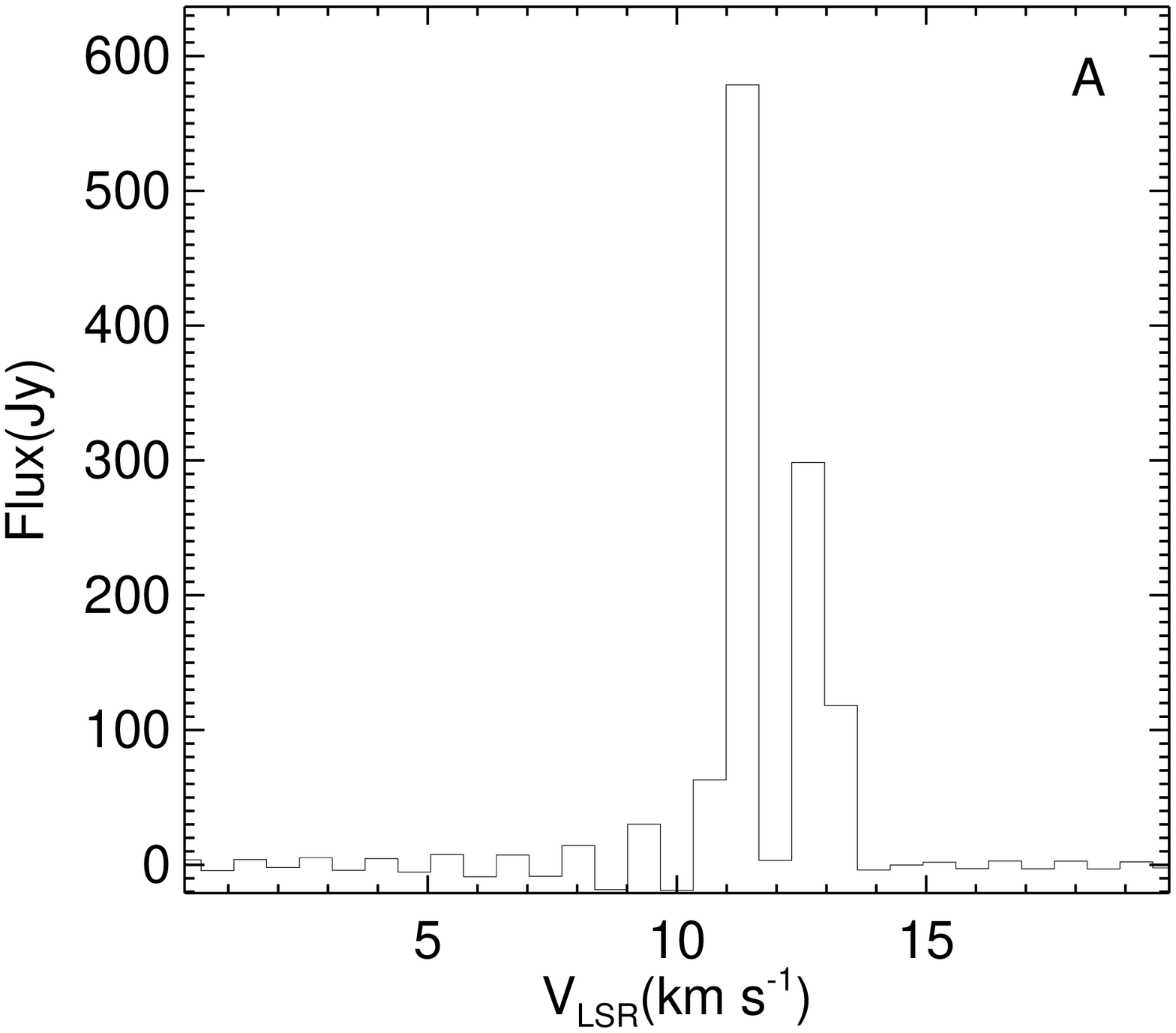}%
\includegraphics[width=5 cm, angle=0]{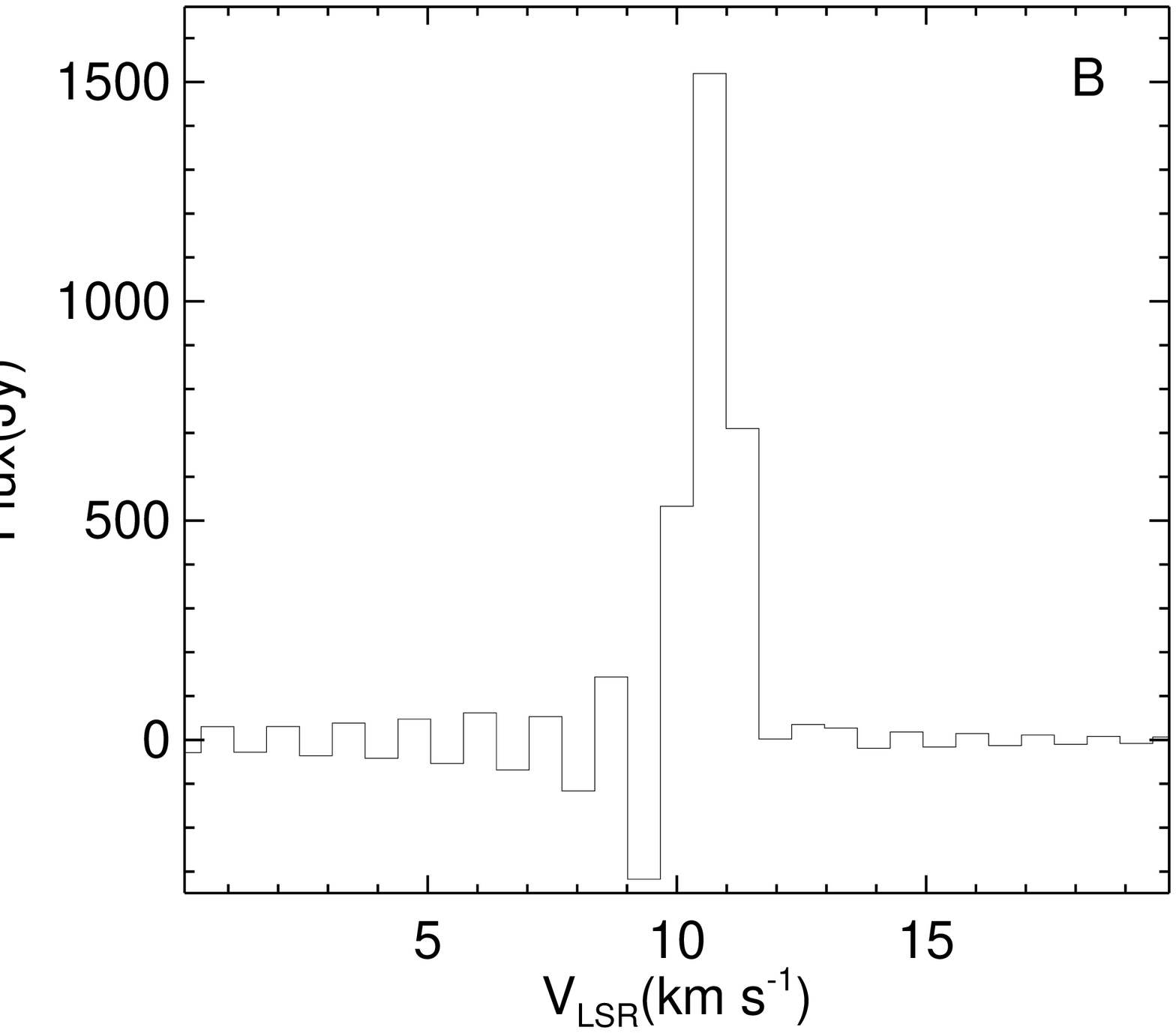}%
\includegraphics[width=5 cm, angle=0]{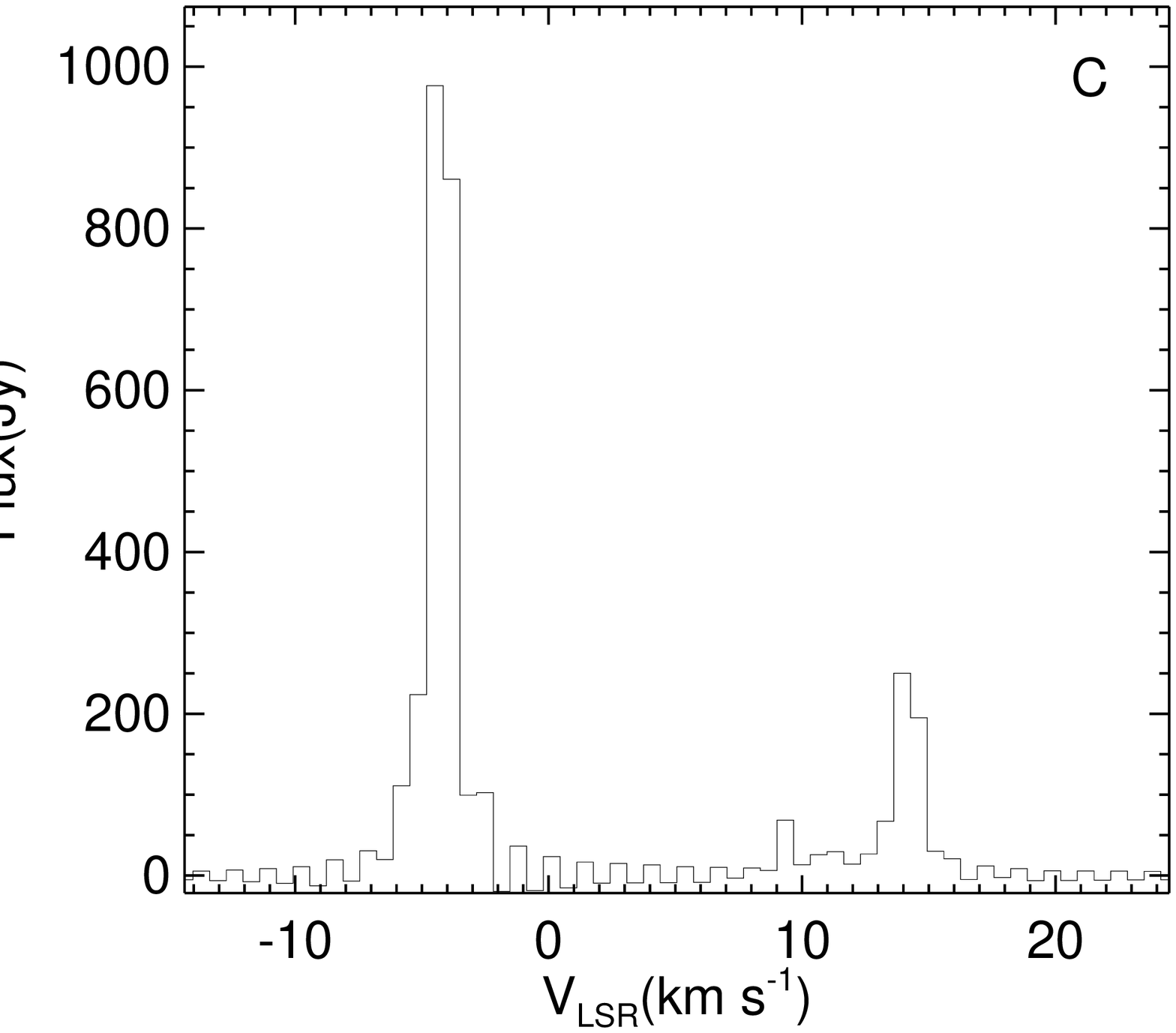}%
                                           
\includegraphics[width=5 cm, angle=0]{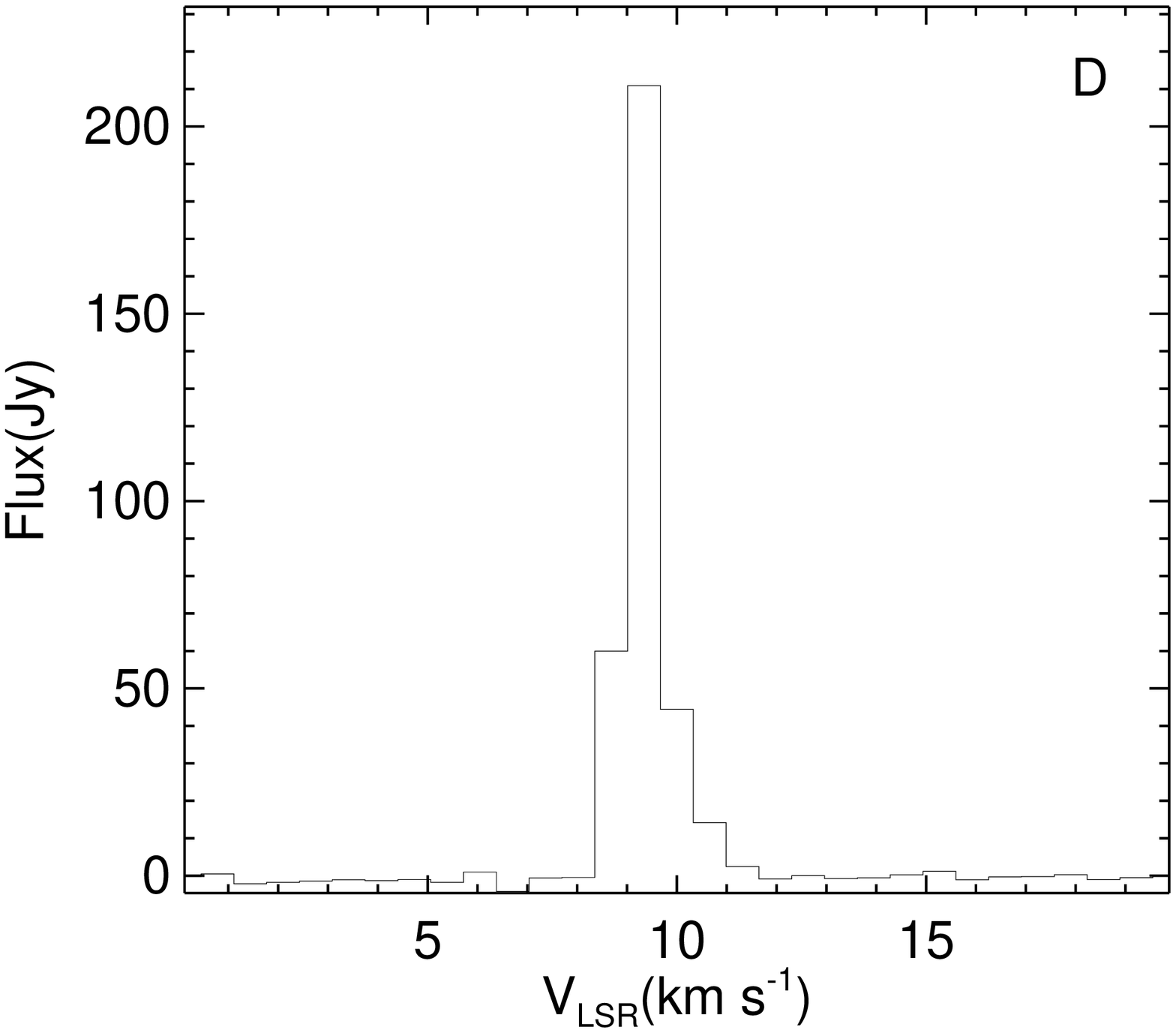}%
\includegraphics[width=5 cm, angle=0]{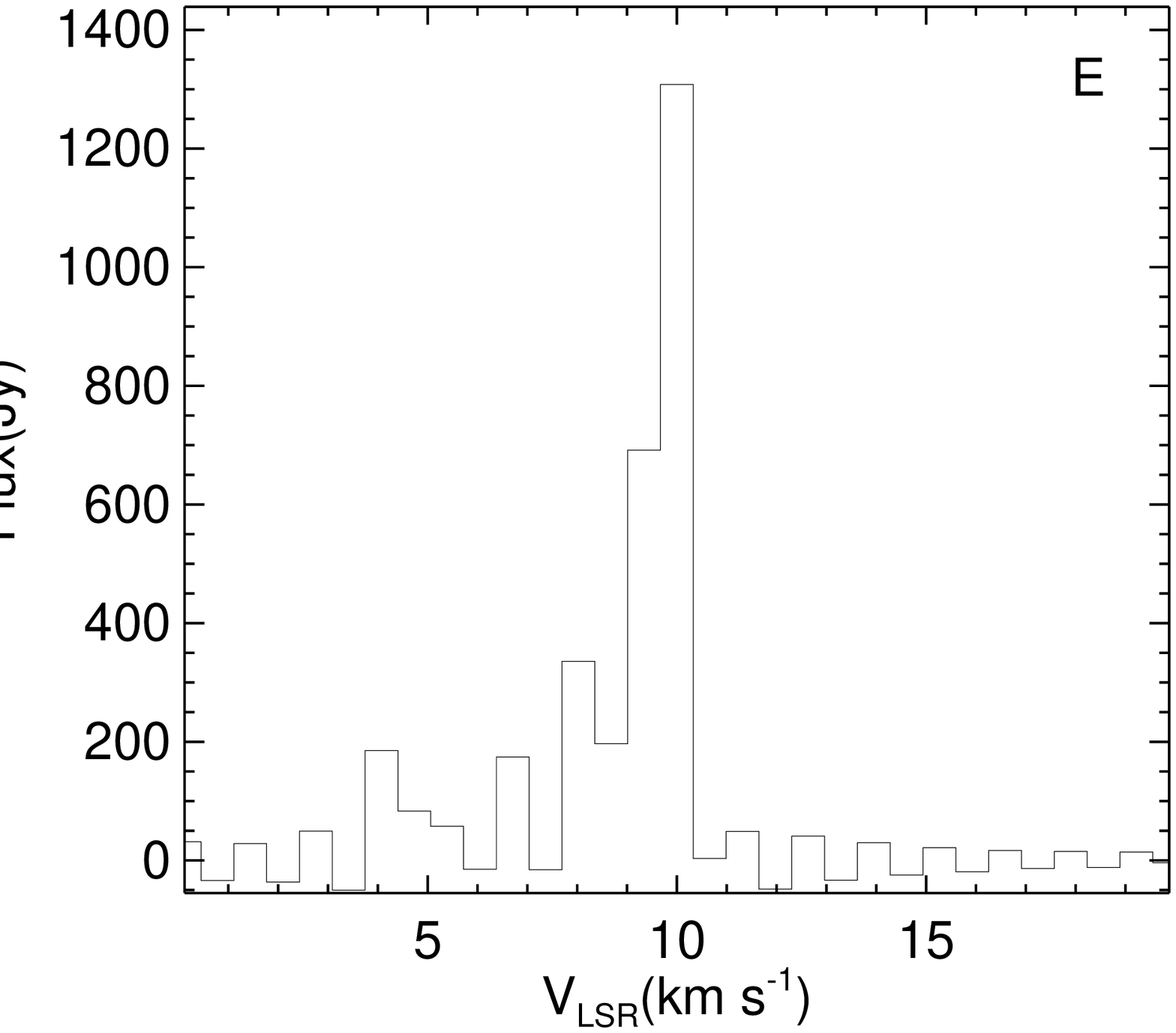}%
\includegraphics[width=5 cm, angle=0]{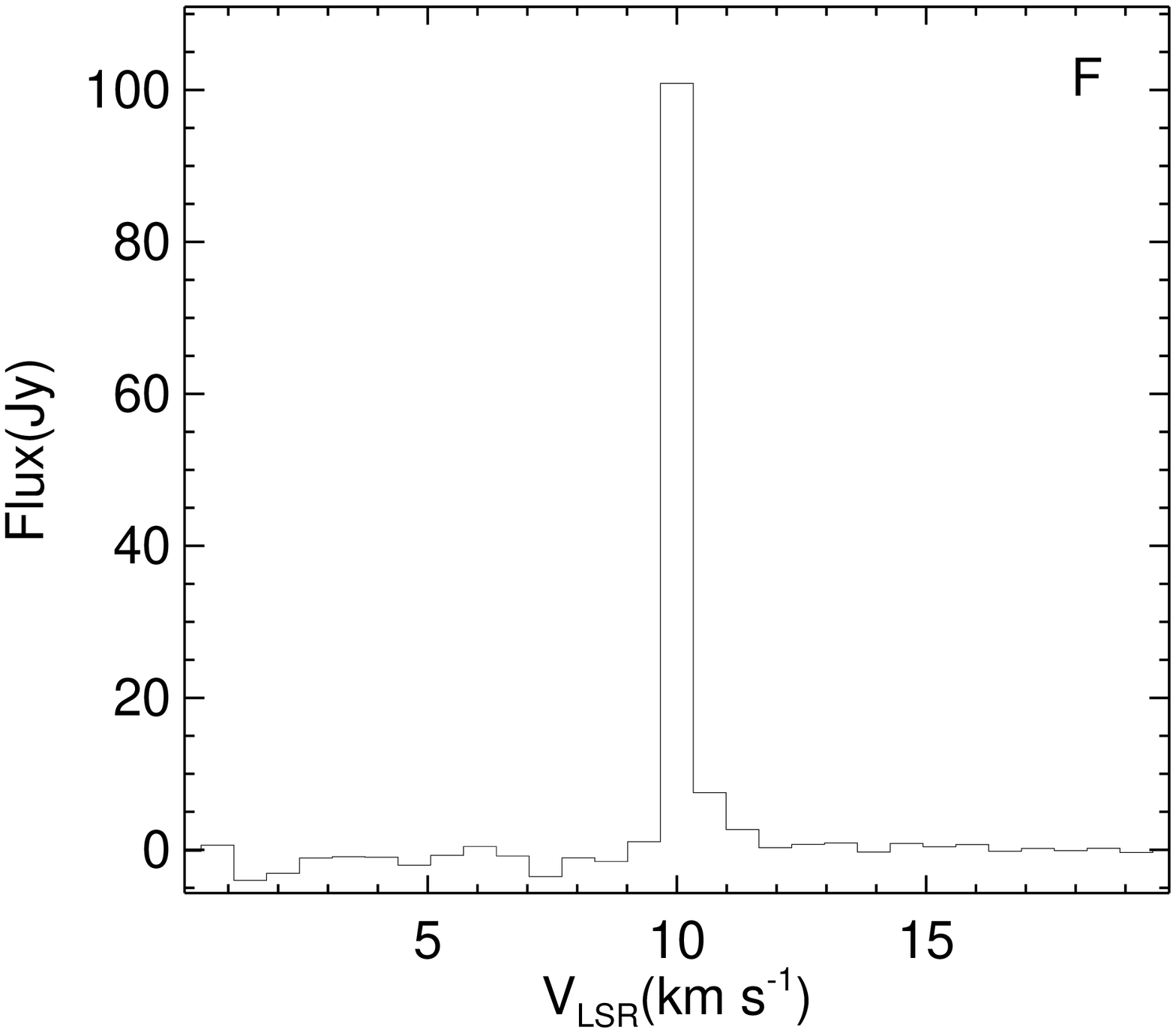}%

\caption{22 GHz H$_2$O maser spectra extracted from the VLA image cube. The six
spectra were extracted from the peak positions labeled A, B, C, D, E and F in Figure
12, by Hanning smoothing the data in spatial coordinates (with a kernel of FWHM $1^{\prime\prime}$). Some spectra show ringing artifacts associated with the
Gibbs phenomenon; these can occur when a strong maser feature is narrower than the channel
width.\label{fig-spectra}} 
\end{figure}

\end{document}